\documentclass[fleqn,usenatbib]{mnras}

\usepackage[T1]{fontenc}
\DeclareRobustCommand{\VAN}[3]{#2}
\let\VANthebibliography\thebibliography
\def\thebibliography{\DeclareRobustCommand{\VAN}[3]{##3}\VANthebibliography}

\usepackage{graphicx}
\usepackage{enumerate}
\usepackage{amssymb, amsmath, amsfonts}
\usepackage{gensymb}
\usepackage{mathtools}
\usepackage{natbib}
\usepackage{color}
\usepackage{hyperref}
\usepackage{ulem}
\usepackage{soul}
\usepackage{xspace}
\usepackage{xcolor}
\usepackage{comment}
\usepackage[pagewise]{lineno}
\usepackage{cellspace}
\newcommand{\CHARISpipeline}{{pyKLIP-CHARIS Post-Processing Pipeline}\xspace}

\newcommand{\HST}{{HST}\xspace}

\newcommand{\CHARIS}{{CHARIS}\xspace}
\newcommand{\SCExAO}{{SCExAO}\xspace}
\newcommand{\pyklip}{{pyKLIP}\xspace}
\newcommand{\klip}{{KLIP}\xspace}
\newcommand{\klipfm}{{KLIP-FM}\xspace}
\newcommand{\wlowrank}{{Weighted Low-rank}\xspace}
\newcommand{\empca}{{EMPCA}\xspace}
\newcommand{\orvara}{{orvara\xspace}}
\newcommand{\daophot}{{DAOPHOT}\xspace}
\newcommand{\GPI}{{GPI}\xspace}
\newcommand{\GRAVITY}{{GRAVITY}\xspace}
\newcommand{\lenslet}{{lenslet}\xspace}
\newcommand{\lenslets}{{lenslets}\xspace}
\newcommand{\Msun}{\mbox{$M_{\sun}$}}

\newcommand{\mas}{\hbox{mas}}

\usepackage{newtxtext,newtxmath}

\title[]{Post-Processing CHARIS Integral Field Spectrograph Data with pyKLIP}

\author[M.~Chen et al.]{Minghan Chen$^{1}$\thanks{E-mail: minghan@physics.ucsb.edu},
Jason J.~Wang$^{2,\; 3}$,
Timothy D.~Brandt$^{1}$,
Thayne Currie$^{4,\; 5}$,
Julien Lozi$^{4}$,
\newauthor
Jeffrey Chilcote$^{6}$ and Maria Vincent$^{7}$
\\\\
$^{1}$Department of Physics, University of California, Santa Barbara, Santa Barbara, CA 93106, United States\\
$^{2}$Department of Astronomy, California Institute of Technology, Pasadena, CA 91125, United States\\
$^{3}$Center for Interdisciplinary Exploration and Research in Astrophysics (CIERA) and Department of Physics and Astronomy,
\\
Northwestern University, Evanston, IL 60208, United States\\
$^{4}$Subaru Telescope, National Astronomical Observatory of Japan, National Institutes of Natural Sciences (NINS), 650 North A`oh\={o}k\={u} Place, \\
Hilo, HI 96720, United States\\
$^{5}$Department of Physics and Astronomy, University of Texas-San Antonio, 1 UTSA Circle, San Antonio, TX 78249, United States\\
$^{6}$Department of Physics and Astronomy, University of Notre Dame, Notre Dame, IN 46556, United States\\
$^{7}$Institute for Astronomy, University of Hawai`i at Mānoa, 2680 Woodlawn Dr, Honolulu, HI 96822, United States
}

\date{Accepted XXX. Received YYY; in original form ZZZ}
\pubyear{2023}

\begin{document}
\label{firstpage}
\pagerange{\pageref{firstpage}--\pageref{lastpage}}
\maketitle

\begin{abstract}

We present the \CHARISpipeline, a Python library that reduces high contrast imaging data for the CHARIS integral field spectrograph used with the SCExAO project on the Subaru Telescope. The pipeline is a part of the \pyklip package, a Python library dedicated to the reduction of direct imaging data of exoplanets, brown dwarfs and disks. 
For PSF subtraction, the \CHARISpipeline relies on the core algorithms implemented in \pyklip but uses image registration and calibrations that are unique to CHARIS.  We describe the pipeline procedures, calibration results, and capabilities in processing imaging data acquired via the angular differential imaging (ADI) and Spectral Differential Imaging (SDI) observing techniques. We showcase its performance on extracting spectra of injected synthetic point sources as well as compare the extracted spectra from real datasets on HD~33632 and HR~8799 to results in the literature. The pipeline is a python-based complement to the SCExAO project supported, widely used (and currently IDL-based) CHARIS Data Post-processing Pipeline (CHARIS DPP) and provides an additional approach to reducing CHARIS data and extracting calibrated planet spectra.

\end{abstract}


\section{Introduction} \label{sec:intro}
Direct imaging is a powerful method to discover extrasolar planets and brown dwarfs and characterize their atmospheres and orbits \citep{Currie2023PPVII}. Planet and brown dwarf companions typically lie at small, sub-arcsecond angular separations: imaging them thus requires very high spatial resolution instruments.  Thermal emission from these companions is also typically far fainter than that of their host stars -- e.g. $\sim$10$^{-4}$--10$^{-7}$ in the near-infrared (IR) for young brown dwarfs and superjovian planets \citep[e.g.][]{Baraffe2003,Spiegel2012}.   Companion contrasts in reflected light can be orders of magnitude steeper.

Furthermore, imperfections and deformations in the optics produces quasi-static speckles in the diffraction pattern that cannot be suppressed through increasing integration time \citep{Marois_2000, Marois_2005, Masciadri_2005}. To tackle these challenges, the field of exoplanet science has seen exciting advancements in instrumentation such as adaptive optics (AO), coronagraphs, integral field spectrographs (IFS) etc., as well as in observing techniques and post-processing algorithms. 

The limiting factor in planet detection is the quasi-static speckle noise. Two observing techniques are often used to separate the speckles from any potential astrophysical signal: Angular Differential Imaging (ADI) \citep{Marois_2006_ADI} and Spectral Differential Imaging (SDI) \citep{Marois_2006_SSDI}. ADI is a technique to take sequences of exposures without tracking the rotation of the field of view (FOV), while keeping the telescope optics aligned. 
An astrophysical signal in such a sequence would move through the FOV relative to the speckle pattern, allowing post-processing algorithms to separate it from the background noise. SDI works on the same principle but along the spectral dimension, where the speckle pattern expands as wavelength increases while an astrophysical signal would remain stationary.  
Some commonly used algorithms developed over the years include Locally Optimized Combination of Images (LOCI; \citealt{Marois_2010_LOCI}), Karhunen-Lo\`eve Image Projection (\klip; \citealt{Soummer2012, Pueyo2015}), Local Decomposition into Low-rank, Sparse, and Gaussian Noise Components (LLSG; \citealt{Gonzalez_2016_LLSG}), etc.

The Coronagraphic High Angular Resolution Imaging Spectrograph ({\CHARIS}) is an integral field spectrograph in the near-infrared for the Subaru Telescope \citep{CHARIS_Design, Groff_2014_CHARIS_contruction}.
\CHARIS uses a lenslet-based design to sample the image plane \citep{CHARIS_Design} and simultaneously obtain a spectrum for each spatial element in a two dimensional (2-D) field of view. This type of design was first implemented on the TIGER IFS \citep{TIGER}. This type of full-field spectrograph makes it possible to extract the spectrum of an exoplanet/brown dwarf companion anywhere in the FOV, enabling the characterization of the companion's properties such as temperature, chemistry, and gravity \citep{Barman_2011, Hinkley_2013, Konopacky_2013, Currie_2014, Currie_2018_kappa,Currie_2023, McElwain_2007}. \CHARIS is also paired with the adaptive optics systems AO188 \citep{AO188} and the Subaru Coronagraphic Extreme Adaptive Optics ({\sl SCExAO}) instrument \citep{Jovanovic_2015} to achieve diffraction-limited resolution. \CHARIS has a low spectral resolution broadband mode that covers J, H and Ks with 22 evenly spaced wavelength bins in logarithmic scale, as well as high-resolution narrowband modes in J, H, or Ks, with 14 wavelength bins in a single narrowband. 

Thus far, all \CHARIS science publications have used the cube extraction pipeline from \citet{Brandt_2017_CHARISPipeline} -- the \CHARIS Data Reduction Pipeline (\CHARIS DRP) -- to convert raw \CHARIS data into data cubes consisting of 2D images at different wavelength channels.   For processing subsequent to this initial step -- e.g. image registration, points spread function (PSF) subtraction and spectrophotometric calibration -- nearly all \CHARIS studies have used the IDL-based \CHARIS Data Processing Pipeline (\CHARIS DPP) \citep{Currie_2020SPIE}\footnote{A Python version of this pipeline should be released sometime in 2023.}.   The \CHARIS DPP pipeline utilizes a variety of algorithms -- KLIP and A-LOCI used in combination with ADI, SDI, and reference star differential imaging (RDI) -- to model and subtract the stellar PSF, yielding the spectra of known planets, brown dwarfs, and disks and producing new discoveries \citep[e.g.][]{Currie_2018_kappa,Goebel_2018, Lawson2020,Steiger_2021_HIP109427,Lawson2020,Lawson2021,Kuzuhara_2022,Swimmer2022,Currie_2022a,Swimmer2022, Currie_2023}.   A separate pipeline infrastructure provides an independent check on the detection of companions and their spectral properties \citep[e.g.][]{Macintosh_2015,Keppler_2018} and widens CHARIS's user base.

In this paper, we introduce a \CHARIS post-processing pipeline under the framework of the \pyklip package, a widely used open source python library for direct imaging \citep{pyKLIP_2015}. \pyklip's PSF modeling is primarily based on the \klip algorithm utilizing principal components analysis (PCA) \citep{Soummer2012}, with an additional PCA-based algorithm, \textit{expectation maximization PCA}, which we add to the existing framework and present in this paper. This pipeline offers an Python-based option for observers to analyze \CHARIS data within a widely-used package framework. The scope of this paper will be limited to processing point sources with ADI and SDI.

We provide an overview of the \CHARISpipeline in Section \ref{sec: pipeline overview}. We explain the detailed astrometric calibrations and provide the calibrated plate scales and north pointing in Section \ref{sec: Astro Cal}. Sections \ref{sec:centroid}, \ref{sec: PSF Modeling}, \ref{sec: spectra extraction} and \ref{sec: specphot cal} describe the post-processing steps and algorithms implemented in the pipeline. We then demonstrate the final calibrated spectral products of the pipeline and offer comparisons with spectra of well known sources in the literature and with injected synthetic sources in Section \ref{sec: full reductions}. Finally, we summarize our work in this paper and discuss potential future updates to the pipeline in Section \ref{sec: Summary}.

\section{Pipeline Overview} \label{sec: pipeline overview}
The \CHARISpipeline is part of the \pyklip package \citep{pyKLIP_2015}, a python library for direct imaging of exoplanets and disks. \pyklip supports multiple instruments and also has the ability to handle generic data products with appropriate formats. While \pyklip was able to reduce \CHARIS data using the generic data mode, it is cumbersome and time-consuming to manually work out the interfacing and calibrations for \CHARIS data. The pipeline we present in this paper fully integrates instrument support into the package with improved image registration and instrument calibrations. We also implement an alternative \wlowrank PSF subtraction method based on the expectation maximization PCA (\empca) algorithm \citep{Bailey_2012}, available for all supported instruments. We comprehensively test and demonstrate the pipeline's performance and show that it produces consistent results with injected sources as well as known real sources in the literature.

\begin{figure*}
    \centering
    \includegraphics[width=0.65\textwidth]{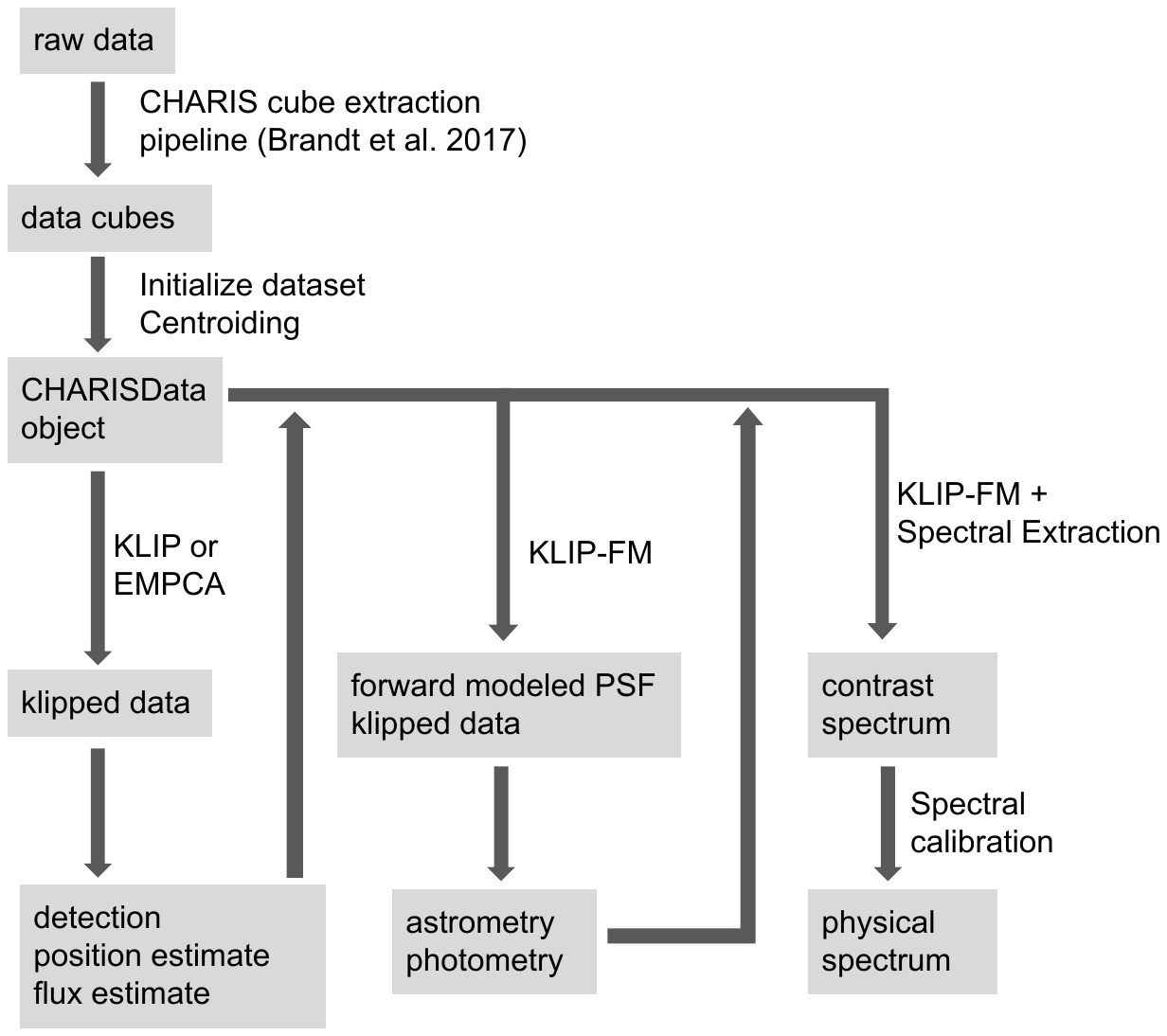} \\
    \vskip -0.25\textwidth
    \caption{\CHARISpipeline flowchart.}
    \label{fig: flowchart}
\end{figure*}

The basic pipeline flowchart is shown in Figure \ref{fig: flowchart}. The \CHARISpipeline takes as inputs the extracted data cubes produced from the \CHARIS raw data reduction pipeline \citep{Brandt_2017_CHARISPipeline}. These are wavelength-calibrated 3-D data cubes with two spatial dimensions and one spectral dimension. The first step is to measure the satellite spot positions and triangulate the stellar position for each image taken at each wavelength. This is done automatically after the user passes in a list of extracted data cubes. Then, one typically first runs \klip or \empca reductions over the full frame (still optimizing over each small section) with a few sets of trial parameters to get a detection. If any companion is detected, one can then run \klip with forward modeling on a single section that contains the companion, which then allows one to fit for the astrometry, photometry as well as extract the spectrum using the \klipfm forward modeling technique \citep{FM_Pueyo_2016}. A detailed documentation of the \CHARISpipeline module and a walk-through of an example \CHARIS dataset can be found on the \pyklip documentation website.\footnote{\url{https://pyklip.readthedocs.io/en/latest/instruments/CHARIS.html}}

\section{Distortion Correction} \label{sec: Astro Cal}

\begin{table*}
    \centering
    \begin{tabular}{ccccc} \hline
    \HST Obs Date & \HST Filter & \HST Detector & Proposal ID & Total Exp Time (s)\\
    \hline
    1997-07-26 & F555W & WFPC2/PC & 6607 & 63 \\
    2000-05-06 & F555W & WFPC2/PC & 8118 & 96 \\
    2006-03-13 & F814W & ACS/WFC & 10775 & 567 \\
    \hline
    \hline
    \CHARIS Obs Date & \# \CHARIS Cubes & Total Exp Time (s) & \multicolumn{2}{c}{\HST Reference Images ID} \\
    \hline
        2017-03-13 & 35 & 2100 & \multicolumn{2}{c}{6607 \& 8118} \\
        2017-04-09 & 4 & 240 & \multicolumn{2}{c}{10775} \\
        2017-09-09 & 20 & 600 & \multicolumn{2}{c}{6607 \& 8118} \\
        2019-03-21 & 2 & 240 & \multicolumn{2}{c}{10775} \\
    \hline
    \end{tabular}
    \caption{HST reference data summary and pairing of HST-CHARIS images used for calibration.}
    \label{table:astrocal data}
\end{table*}

\subsection{Plate Scales and Position Angle Corrections} \label{subsec: astrocal}

Distortion correction is the mapping between coordinates in a distorted image to the coordinates in a distortion-free image. Measuring the distortion correction in high contrast imaging data usually only involves calibrating the plate scale (potentially different along $x$ and $y$ axes) and the position angle (PA) zero point of the instrument. The PA zero point is the direction of true north in the telescope's field of view. For a high contrast imaging instrument, visual binaries with well-known separations can be used as distortion-free references to calibrate the plate scale and north pointing. Higher order distortion effects are usually negligible due to the small fields of view of high contrast imaging instruments. For \CHARIS, the lenslets disperse $134 \times 134$ microspectra onto the detector, giving us a field of view of roughly $2.2 \arcsec \times 2.2 \arcsec$. 

\begin{figure}
    \centering
    \includegraphics[width=0.48\textwidth]{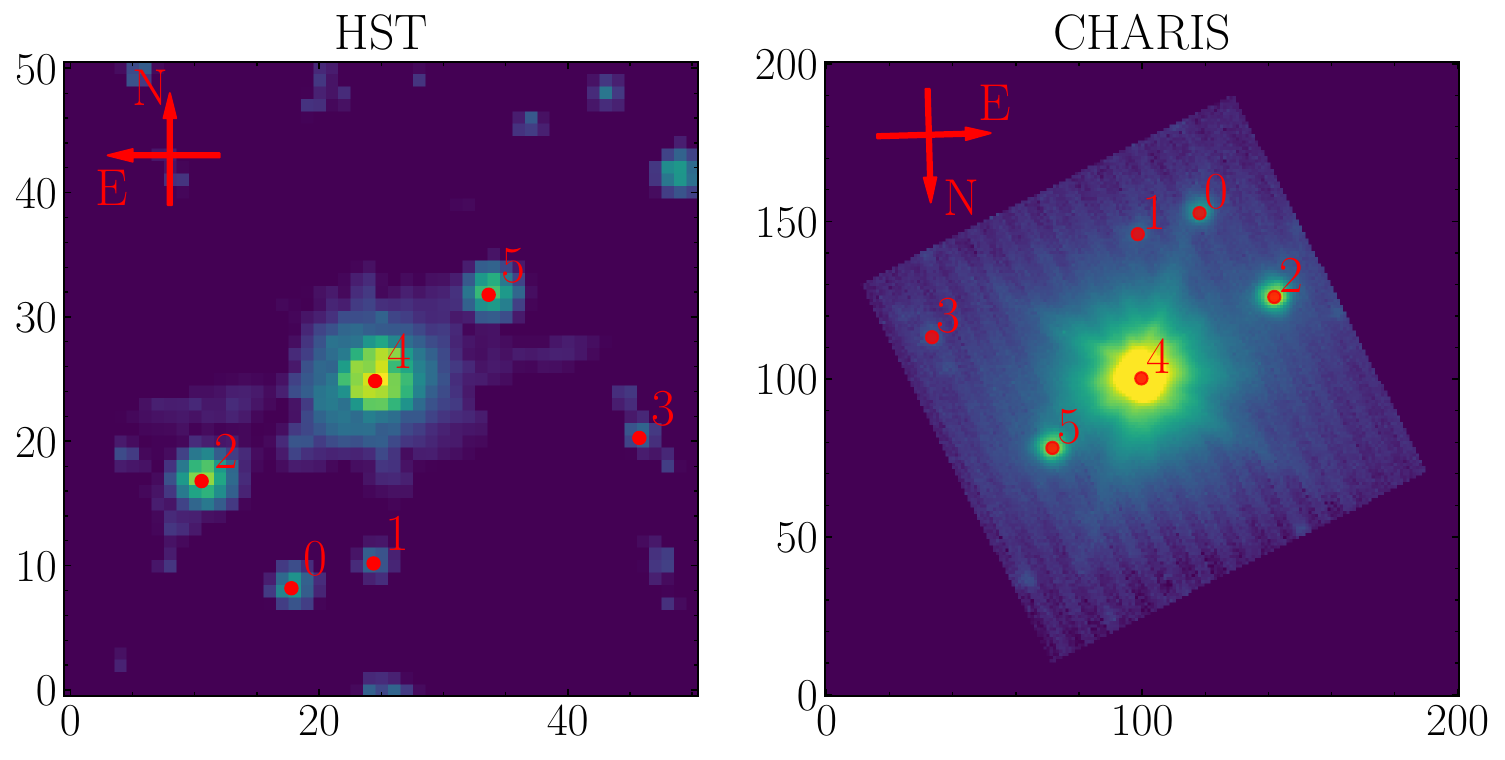}
    \caption{An example of the \HST and \CHARIS calibration data. Annotated in red are the \daophot extracted positions of the point sources that are used for plate scale calibration for this particular example.}
    \label{fig: platecal images}
\end{figure}

In this section, we use multiple datasets taken at different times to calibrate the plate scale and north pointing for \CHARIS, as well as examine the stability of these parameters over time. We first use a globular cluster to measure the plate scales along the x and y axes separately. This is possible because there are multiple stars in the field of view with different separation vectors, allowing us to constrain the plate scale along x and y axes separately. The drawbacks compared to using a bright visual binary include worse spatial resolution in the reference images (\HST images) of the star cluster, and relatively low SNR and Strehl ratio of the stars in CHARIS images, both lead to larger uncertainties in the positions of the reference stars used for calibration. The small field of view of CHARIS can only capture a handful of stars in the cluster (see Figure \ref{fig: platecal images} as an example). The advantage of having multiple stars compared to a binary makes up for the aforementioned drawbacks, such that the constraints on the zeroth and first order distortion coefficients have comparable error bars as those from a binary calibration. However, a few stars is not enough to provide meaningful constraints for higher order distortion corrections with more coefficients. Therefore, we limit our star cluster calibration to only first order corrections. We then use a visual binary to measure an additional calibration result at a different epoch. 

\subsubsection{Calibration Using the M5 Star Cluster} \label{subsubsec: HST astrocal}
We express the transformation between \lenslet coordinates in a \CHARIS image and the corresponding distortion free coordinates as follows:
\begin{equation} \label{eqn:platecal}
    \begin{bmatrix}
    x_{\rm cal} \\
    y_{\rm cal} \\
    \end{bmatrix} = \begin{bmatrix}
    x_0 \\
    y_0 \\
    \end{bmatrix} + \mathbf{R} \begin{bmatrix}
    S_x\;x \\
    S_y\;y \\
    \end{bmatrix}
\end{equation}
where $(x, y)$ are the \lenslet coordinates in an uncalibrated \CHARIS image, $S_x$ and $S_y$ are the plate scales along the $x$ and $y$ axis of a \CHARIS image, $\mathbf{R}$ is the 2D rotation matrix associated with a counter-clockwise rotation of $\theta$, $(x_{\rm cal}, y_{\rm cal})$ are the calibrated coordinates in a distortion free coordinate frame, and $(x_0, y_0)$ are translation offsets. We formulate the plate scale along two orthogonal directions separately to test if the plate scales along the x and y axes of the detector are indeed consistent. 

We use images of the M5 star cluster taken by \CHARIS and \HST for our calibration under the formalism of Equation \eqref{eqn:platecal}. We find the most ideal \HST images available for this task to be the ones taken by either the Planetary Camera (PC) of WFPC2, or the Wide Field Channel (WFC) installed on the Advanced Camera for Surveys (ACS), due to their having high resolution, high quality images with few saturated stars. We use the calibrated \HST images available on the Hubble Legacy Archive that were taken via these two channels and contain the stars imaged by CHARIS. The images we selected come from \HST proposal IDs 6607 (PI Ferraro), 8118 (PI Piotto), and 10775 (PI Sarajedini), listed in Table \ref{table:astrocal data}. CHARIS observations have 5 different small fields of view over four epochs, while the HST images cover much larger fields of views. The stars imaged by CHARIS appear in one or two of the three HST images, but not all of them. The pairing of CHARIS and HST images used for calibration is also listed in Table \ref{table:astrocal data}. The WFPC2 data was reduced using the OPUS pipeline \citep{OPUS_pipeline} and the ACS/WFC data was reduced using the calibration pipeline CALACS \footnote{\url{https://www.stsci.edu/hst/instrumentation/acs/software-tools/calibration-tools}}. All HST images we selected have a plate scale of $50$ \mas/pixel and have the y-axes oriented towards north, after the pipeline calibrations. For ACS/WFC, the relative precision of the pixel scale and the orientation are both on the order of $\sim 6\times10^{-6}$ \citep{2007_acs_calibration}, which is negligible compared to CHARIS and won't be factored into our error propagation. For WFPC2/PC, the precision of the the orientation is $\sim 0.03\degree$ \citep{Holtzman_1995_WFPC2}, which is factored into our error propagation.

To extract the positions of the stars in HST images, we follow the iterative effective PSF (ePSF) building and fitting approach described in \citep{Anderson_2000, Anderson_2006}, which is implemented in the Photutils python package \citep{Stetson_1987, photutils110}. For CHARIS images, there are not enough stars within the field of view to build ePSFs, and the seeing is also variable over time. Therefore, we fit for a 2D Moffat profile to each star in every halo-subtracted exposure. We match the stars that are present in both \CHARIS and \HST images, then use the positions of the stars to fit for \{$x_0$, $y_0$, $\theta$, $S_x$, $S_y$\} as given in Equation \eqref{eqn:platecal}. In this case, $S_x$ and $S_y$ are the scaling factors that convert the \CHARIS plate scales along each axis to the known \HST plate scale. The angle $\theta$ is the counter-clockwise angle a \CHARIS image needs to be rotated to align with an \HST image whose y-axis is already aligned with celestial north. Thus, $-\theta$ is equivalent to the PA of the celestial north in a \CHARIS image. The parallactic angle, $\eta$, is recorded for every \CHARIS image at the time of exposure. It provides an estimate for the angle $\theta$: $\theta_{\rm est}=\eta + 113\degree$, where $113\degree$ is the wavelength dispersion angle of the lenslets on the detector. During post-processing, CHARIS images are rotated counter-clockwise by $\theta_{\rm est}$ to align the y-axis with the celestial north. However, while the dispersion angle is known and very stable, $\eta$ may have some offset relative to the true parallactic angle. This offset may even vary slightly from observation to observation due to small changes in optics alignment. By fitting for the angle $\theta$, we measure the north pointing correction to the default estimate: 
\begin{equation}
\theta_{\rm north} = \theta_{\rm est} - \theta
\end{equation}
This $\theta_{\rm north}$ is the PA zero point in a CHARIS image aligned via the default estimate.

The \CHARIS field of view is small ($2.2 \arcsec \times 2.2 \arcsec$) and can capture only a handful of stars of the M5 star cluster. Between 2017 and 2020, CHARIS took several sequences of the M5 cluster, each sequence capturing a small section of the cluster. We select by eye the \CHARIS observations with at least $3$ stars within the field of view that are not within 5 lenslet of the boundary. And we require that at least one pair of stars are separated by more than half an arcsecond, because small clumps of closely separated stars would produce large uncertainties and biases. We use a total of 61 \CHARIS data cubes of M5, observed on 4 different dates with 5 different fields of view that are also imaged by either PC or WFC. A \CHARIS data cube consists of images at different wavelengths. We collapse each \CHARIS data cube into a single image by taking the mean of the data cube along the wavelength dimension, while discarding the wavelength slices with significant atmospheric absorption (two wavelength channels at $1.37 \mu m$ and $1.87 \mu m$ for CHARIS JHK broadband are discarded). We then use the collapsed \CHARIS images for calibration analysis.

We use the emcee package \citep{emcee_2013} to perform MCMC fits of the parameters \{$x_0$, $y_0$, $\theta$, $S_x$, $S_y$\} in equation \ref{eqn:platecal}, where the log-likelihood function is given by:
\begin{equation}
    \label{eqn:platecal lnlike}
    ln\mathcal{L} = -\frac{1}{2}\sqrt{\sum_{i}^{N_{stars}}\frac{|\mathbf{x}^{cal}_{i} - \mathbf{x}^{HST}_{i}|^{2}}{\sigma^{2}_{cal} + \sigma^{2}_{HST}}}\\
\end{equation}
where $\mathbf{x}^{cal}_{i}$ is the calibrated coordinate vector of the $i$th star, calculated from equation\ref{eqn:platecal}, and $\mathbf{x}^{HST}_{i}$ is the \daophot extracted position of the same star in the \HST image.

\subsubsection{Calibration Using Visual Binaries}
The results of the previous section are consistent with a single plate scale across the field of view, as shown in Section \ref{subsec: astrocal results}. Therefore, we also carry out the calibration of plate scale and PA under the assumption of a single consistent plate scale, using images of a visual binary with a well-known separation. In this case, the calibration can be expressed very simply as:
\begin{align}
S &= \frac{\rho_{\rm cal}}{\Delta_{\rm lens}} \label{eqn: binary cal pscale}\\
\theta_{\rm north} &= \theta_{\rm CHARIS} - \theta_{\rm cal} \label{eqn: binary cal PA}
\end{align}
where $\rho_{\rm cal}$ is the known separation of the binary in angular units. $\Delta_{\rm lens}$ is the measured separation from \CHARIS images in \lenslets. $S$ is the plate scale in ${\rm mas}/{\rm \lenslet}$. $\theta_{\rm CHARIS}$ is the PA of the binary companion in \CHARIS images using the default north pointing set by the data reduction pipeline. $\theta_{\rm cal}$ is the known PA of the binary. And $\theta_{north}$ is the PA of the true north in \CHARIS images. 
We carried out this type of calibration using two visual binary systems: HIP 55507 and HD 1160. We describe the two datasets in the following two subsections and show our results alongside previous binary calibrations in other works \citep{Kuzuhara_2022, Currie_2018_kappa, Currie_2020_HD33632, Currie_2022a} in Section \ref{subsec: astrocal results}.

\subsubsection{HIP 55507}
We have a recent \CHARIS dataset of HIP~55507, a $\approx$0$.\!\!''75$ visual binary. This binary was observed by \CHARIS on UT 2022 February 27th as an astrometric calibrator for the characterization of the newly discovered T-Dwarf companion, HIP 21152 B \citep{Franson_2023}. HIP~55507 was previously imaged multiple times by Keck/NIRC2, on UT 2012 January 07 and 2015 May 29 (PI Justin Crepp) as part of the TRENDS survey \citep{Gonzales_2020} and also on UT 2021 December 21st (PI Kyle Franson). The calibrated NIRC2 images has an uncertainty in the north position angle of $0\degree.02$ and an uncertainty in astrometry of $0.5\,\mas$ \citep{Service_2016_NIRC2_Distortion_Solution}. The Keck/NIRC2 astrometry of this companion was precisely measured by \cite{Franson_2023} to be $759 \pm 5$ mas. The non-negligible orbital motion between Keck/NIRC2 and \CHARIS observations are accounted for via an orbital fit carried out by the orbit fitting code, Orbits from Radial Velocity, Absolute, and/or Relative Astrometry (\orvara), \citep{orvara_2021}. The separation of this binary in \CHARIS images is measured by fitting a 2D Airy disk to the companion, HIP~55507 B, in aligned, background subtracted, and exposure-averaged \CHARIS images at different wavelengths, and then the results are averaged over wavelength. A full PSF subtraction is not necessary for this calibrator due to the high SNR and diffraction limited PSF of the binary companion. We anchor the CHARIS astrometry to the Keck/NIRC2 astrometry and measure the plate scale and truth north using equations \ref{eqn: binary cal pscale} and \ref{eqn: binary cal PA}. The results are presented in Section \ref{subsec: astrocal results}.

\subsubsection{HD 1160}
HD 1160 is a system with two wide (sub)stellar companions \citep{Nielson_2012_HD1160}, one of which (HD~1160 B) is visible within the CHARIS field of view. This system has been used to anchor the astrometric calibration in previous CHARIS papers by comparing CHARIS astrometry of HD 1160 B to that obtained (near-)contemporaneously from the precisely-calibrated Keck/NIRC2 \citep{Kuzuhara_2022, Currie_2018_kappa, Currie_2020_HD33632, Currie_2022a}.  Here we add four subsequent calibrations of CHARIS datasets of HD 1160, observed in 2021 October, 2022 January, 2022 September, and 2022 December (PIs: Thayne Currie, Maria Vincent).   We pin the astromeric calibration of these data to unsaturated Keck/NIRC2 data obtained in the $K_{\rm s}$ and $L_{\rm p}$ filters on October 15, 2021 (pinning Oct 2021 CHARIS astrometry), January 15, 2022 (pinning Jan 2022 CHARIS data), and September 8, 2022 (pinning Sept and Dec 2022 CHARIS data) (PIs: Thayne Currie, Taichi Uyama).   

The CHARIS data were reduced using the CHARIS DPP; the Keck/NIRC2 data were reduced using a general-use, well-tested code \citep{Currie2011}.    In each data set, HD 1160 B is detectable without PSF subtraction techniques, typically at SNR $\sim$ 10-20 with NIRC2 and SNR $\sim$ 35--135 with CHARIS.   
The results are presented in Section \ref{subsec: astrocal results}.

\subsection{Characterizing Precision and Biases}
The average centroid error of the itetrative ePSF fitting for stars in the HST images is $\sim$0.05 pixel, going down to $0.02$ pixel for the well isolated bright stars, and up to $0.08$ pixel for stars with relatively close neighbors. For \CHARIS images, the average centroid error is $\sim 0.7$ \lenslet. In this section, we further examine the systematic errors and potential biases using mock star cluster data. 

For each set of stars within the CHARIS field of view in the real M5 star cluster data, we use the extracted positions of those stars in an \HST image as the ground truths for our mock \HST observation. To obtain the ground truth positions for the corresponding mock \CHARIS observation, we choose a ground truth rotation angle similar to the best fit rotation for the real data, as well as choosing 16\;mas/\lenslet as the true plate scales for x and y axes, and transform the mock \HST true positions into mock \CHARIS true positions by reversing equation \ref{eqn:platecal}. Then, we generate ``measured positions'' of these stars by adding random gaussian errors with $\sigma=0.1$\;pixel or \lenslet to the ground truth positions for both mock \HST and mock \CHARIS observations. We then fit for the parameters in equation \ref{eqn:platecal} for a pair of mock observations and compare it to the ground truths to get a residual. To examine the errors and potential biases, we generate 100 pairs of such mock \HST and \CHARIS observations and look at the average and the scatter of the residuals. Since the precision would depend on the configuration of the calibration stars in the field of view, we perform this investigation separately for each of the 5 fields of view of the real data.

\subsection{Results} \label{subsec: astrocal results}
\begin{figure}
    \centering
    \includegraphics[width=\linewidth]{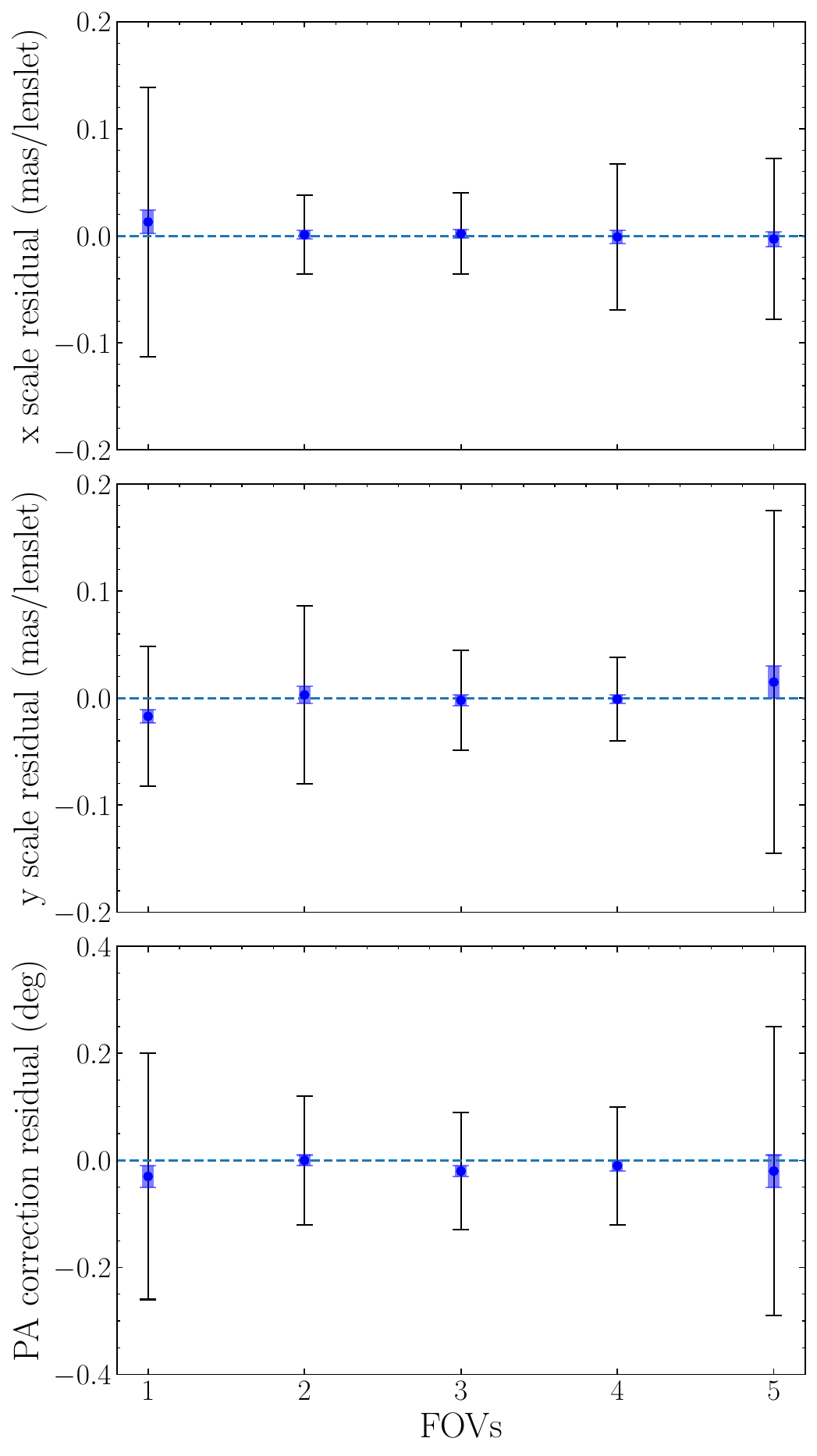}
    \caption{Systematic error estimates for the configurations of the calibration stars used for the different fields of view. Each point is the measured plate scale averaged over all samples for each field of view. The black error bar is the scatter of the measured plate scale of all samples for each field of view. The blue error bar is the error of the mean, which is the black error bar divided by $\sqrt{N_{\rm sample}}$. The blue error bars show that the results are consistent with zero, indicating there is no significant bias in the calibration.}
    \label{fig: platecal systematics}
\end{figure}

The results for the precision and bias examination of the star cluster calibration method are shown in Figure~\ref{fig: platecal systematics}. The data points are the averaged residuals, relative to the ground truth plate scale, over 100 pairs of mock observations for each field of view. The thin black error bars are the RMS scatter of the residuals of the 100 samples in each field of view. The thick blue error bars are the errors on the mean, suggesting that all the residuals are mostly consistent with zero. For every single pair of \HST and \CHARIS images used in the real data calibration, the RMS scatter (black error bars) are used and propagated to get the final errors of the corresponding field of view for the real data. 

\begin{figure}
    \centering
    \includegraphics[width=\linewidth]{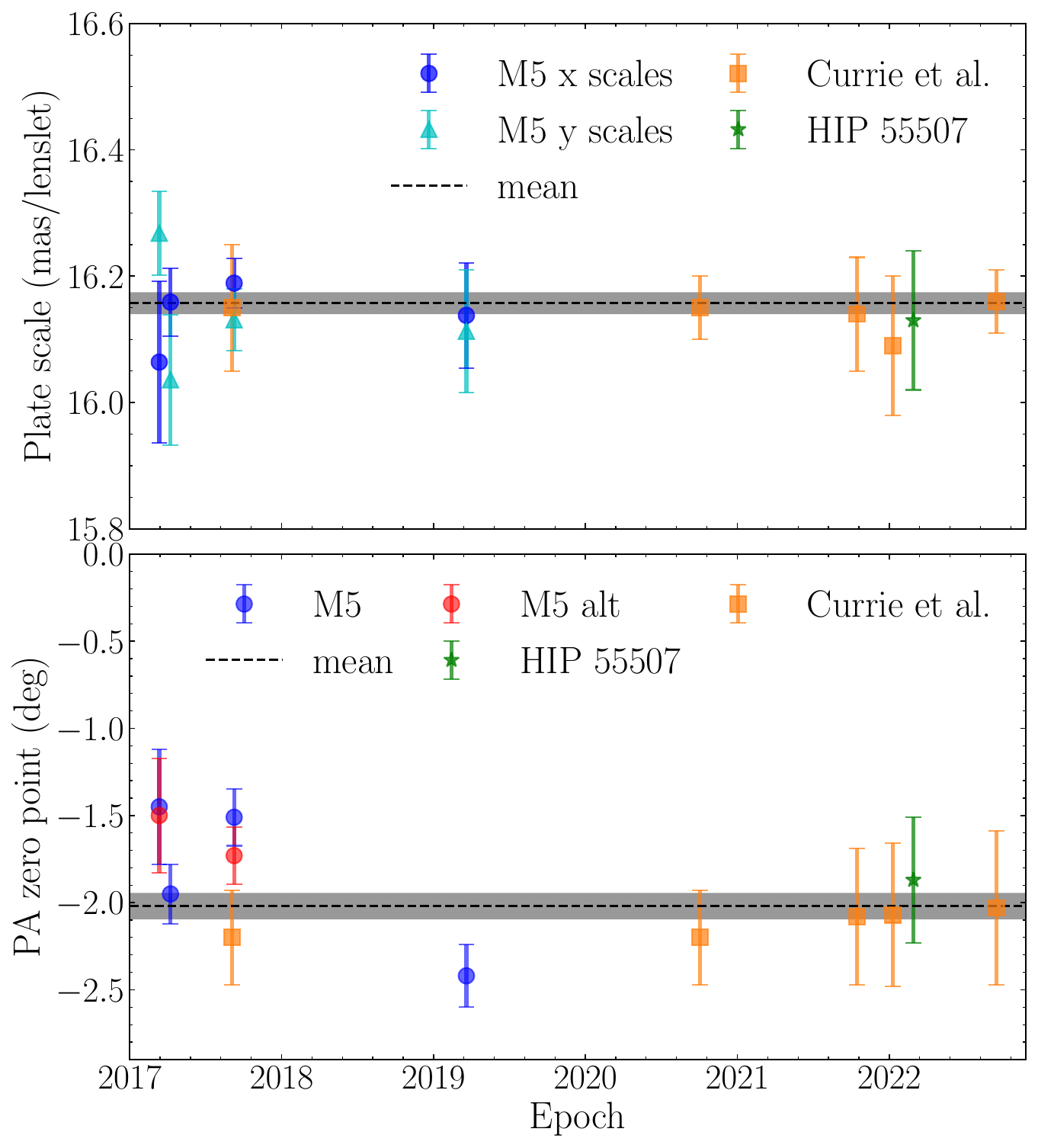}
    \caption{Best fit \CHARIS plate scales and PA zero points. The M5 cluster calibration provides separate plate scales along the x and y axes of the detector that agree with each other to $\sim 1\sigma$, and are also consistent with other measurements with a single plate scale. The red points in the bottom panel show the results of the M5 calibration after excluding a star that may be causing some systematic bias. The dashed lines indicate the inverse variance weighted mean plate scale and PA zero point. The dark shaded regions are the standard errors on the mean. }
    \label{fig: platecal}
\end{figure}

The results of the star cluster calibration and binary calibration of the \CHARIS plate scale and PA zero point are summarized in Table \ref{table:platecal} and Figure \ref{fig: platecal}. The mean values are plotted as the dashed line and the errors on the mean are plotted as the dark shaded regions. The binary calibration results include two measurements (first two in the Binary Calibration section of Table \ref{table:platecal}) from previous work \citep{Currie_2018_kappa, Currie_2020_HD33632, Currie_2022a}, with the caveat that these two results are averaged over several epochs. However, our results from individual epochs suggest that there is not significant variation over time.

The M5 star cluster calibration measures the plate scales with errors on the order of $\sim$0.1 \hbox{mas}.  They show that the plate scales along the $x$ and $y$ axes are consistent with each other to about $1\sigma$. The plate scales are also consistent with the plate scales measured from binary calibrations, and seem to be stable over time. 

The overall average plate scale of all calibrations as well as all calibrations derived from binary companions is $16.15 \pm 0.02$ mas/\lenslet, which is the same as the $16.15 \pm 0.05$ mas/\lenslet value adopted in previous papers and in the current version of the CHARIS DPP.   Measurements from binary calibrations are dominated by uncertainties in the stellar centroid position and the intrinsic SNR of the Keck data to which CHARIS astrometry is pinned.  Future unsaturated, high SNR observations of HD 1160 B and other binaries could improve the CHARIS plate scale precision.

Most PA zero point measurements agree within error bars with the default value of $-2.2 \pm 0.27 \degree$, which has been used to analyze CHARIS datasets so far \citep{Currie_2018_kappa,Currie_2023}. The average position angle offset from all measurements is $-2.03 \pm 0.08 \degree$. From just the binary calibrators, the average offset is $-2.11 \pm 0.13 \degree$.

There are two epochs of M5 calibration data that produced PA zero points that are $\sim 1\sigma$ above the total average. While $1 \sigma$ is still within reasonable statistical uncertainty, it is worth noting that these two epochs of CHARIS data of the M5 cluster share the exact same field of view, and this correlation led us to suspect there might be some systematic bias. We took a closer look at these two epochs by carrying out a Jackknife resampling of the stars in the field of view used for the calibration. Among the Jackknife samples, the exclusion of the same star in both epochs caused the PA zero point to change more significantly than the exclusion of any other star, which brought both epochs into better agreement with the other measurements. A potential cause is that the star in CHARIS images partially overlaps the diffraction spike created by the support spider of the Subaru Telescope, which could induce some bias in the measured centroid of the star. However, this is not sufficient to conclude that there is real bias induced by the star. Therefore, we present both the results with the exclusion of this star (red data points) and the results without excluding any star (blue data points) in the bottom panel of Figure \ref{fig: platecal}. 

Overall, the calibrations suggest that the plate scale and north pointing are stable over time. But it is possible that there could be some small fluctuation of the north pointing over time on the order of the size of the error bars, because the \CHARIS + \SCExAO instrument is mounted on rubber mounting and can be craned in and out of position between observations \citep{Jovanovic_2015_SCExAO, Kuzuhara_2022}. Therefore, we encourage observers to take contemporary data of a binary calibrator with precisely known astrometry in addition to science data, and use the calibrator to measure the PA zero point for their observation. If contemporary calibration data is not available, then we recommend adopting the average results of all binary calibrations compiled in Table \ref{table:platecal} or the average of all binary calibrators.  Using this calibration instead of the default one, we note that the revised astrometry for recently-discovered companions shows inconsequential differences with published values: e.g. a difference of $\approx$0.2 mas in the east and north positions for HIP 99770 b from October 2021 compared to measurement uncertainties of 4 mas \citep{Currie_2023}.

\begin{table*}

\caption{Best fit results for \CHARIS plate scales and PA zero point corrections.}
\label{table:platecal}
\begin{tabular}{cccc}
\multicolumn{4}{c}{M5 Star Cluster Calibration}\\
\hline
CHARIS Obs Date & x scale (mas/\lenslet) & y scale (mas/\lenslet) & PA zero point (deg)\\
\hline
2017-03-13 & $16.06 \pm 0.13$ & $16.27 \pm 0.07$ & $-1.50 \pm 0.33^{b}$ \\
2017-04-09 & $16.16 \pm 0.05$ & $16.04 \pm 0.10$ & $-1.95 \pm 0.17$ \\
2017-09-09 & $16.19 \pm 0.04$ & $16.13 \pm 0.05$ & $-1.73 \pm 0.16^{b}$ \\
2019-03-21 & $16.14 \pm 0.08$ & $16.11 \pm 0.09$ & $-2.42 \pm 0.18$ \\
\hline
\hline
\multicolumn{4}{c}{Binary Calibration$^{a}$}\\
\hline
CHARIS Obs Date & System & Plate scale (mas/\lenslet) & PA zero point (deg)\\
\hline
2017-07-16 & & & \\
2017-09-04 & HD 1160 & $16.15 \pm 0.10$ & $-2.20 \pm 0.27$ \\
2017-09-06 & & & \\
\hline
2018-10-18 & & & \\
2020-10-03 & HD 1160 & $16.15 \pm 0.05$ & $-2.20 \pm 0.27$ \\
2020-12-07 & & & \\
2021-09-11 & & & \\
\hline
2021-10-15 & HD 1160 & $16.14 \pm 0.09$ & $-2.08 \pm 0.39$ \\
\hline 
2022-01-09 & HD 1160 & $16.09 \pm 0.11$ & $-2.07 \pm 0.41$ \\
\hline
2022-09-15 & HD 1160 & $16.16 \pm 0.05$ & $-2.03 \pm 0.44$ \\
\hline
2022-12-29 & HD 1160 & $16.17 \pm 0.11$ & $-2.17 \pm 0.33$ \\
\hline
2022-02-28 & HIP 55507 & $16.13 \pm 0.11$ & $-1.87 \pm 0.36$ \\
\hline
\hline
Binary Calibration Average & & $16.15 \pm 0.02$ & $-2.11 \pm 0.13$\\
\hline
Overall Average & & $16.15 \pm 0.02$ & $-2.03 \pm 0.08$\\
\hline
\end{tabular}\\
{a }{The first two binary calibration results using HD 1160 are from \cite{Currie_2018_kappa} and \cite{Currie_2022a}. They averaged over three and four epochs of CHARIS data, respectively.}\\
{b }{These two PA zero points correspond to the red points in Figure \ref{fig: platecal}.}
\end{table*}

\section{Image Registration} \label{sec:centroid}
The first step in processing coronagraphic data is to accurately determine the position of the centroid of the bright primary star in the images. This allows us to re-register all images through interpolation to align them relative to each other in order to model the diffraction pattern of the primary star over a sequence of images. 

In high contrast coronagraphic imaging, the bright primary star at the center of the image is occulted by a coronagraph to block out most of the star light. To enable precision astrometry for CHARIS, a diffractive grid generated by sine waves on the deformable mirror (DM) of \SCExAO in the pupil plane projects the primary star onto the focal plane with a fixed contrast \citep{Jovanovic_2015, Jovanovic_2015_SCExAO, Sahoo_2020}, before the star is occulted. These projections of the primary star, which we refer to as ``satellite spots'', are much dimmer copies of the primary star's PSF that can be imaged within the dynamic range of the camera. The pattern of the diffractive grid can be varied to control the placement of these satellite spots. The period of the grid controls the separation of the satellite spots from the central star in units of $\lambda / d$. More periods within the pupil size projects the spots further away from the center. The orientation of the sine waves controls the orientation of the satellite spots, and the sine waves can be placed along one direction or two orthogonal directions on the DM to control the number of spots (2 or 4) \citep{Jovanovic_2015}. Fitting for the positions of these satellite spots allows us to triangulate the centroid of the occulted primary star and then align all the images for PSF/speckle modeling.

An early version of the \pyklip-\CHARIS did this with a simple maximum likelihood fit of a 2D PSF model (Airy Function or Gaussian) to every satellite spot, given user specified initial guesses. We refer to this as the "local centroid algorithm" henceforth. This method does reasonably well when initialized close to the correct spot parameters, but is inconsistent at finding the global best-fit and can have large errors for many individual satellite spots. The final centroid of the primary star in each frame is determined by the average position of the spots determined separately at each wavelength; it is thus prone to outliers. We present an alternative global centroid algorithm that fits the centroid of all images in a dataset simultaneously, taking into account the constraint that the spots lie on the vertices of a square, as well as the cross-correlation between images, thus improving the stability and robustness of the image centroids.

\subsection{Global Centroid Algorithm}\label{subsec: global centroid}

We adopt a centroiding approach based on both matching the image slices within a data cube at different wavelengths, and matching different data cubes with one another. Both of these components of our method use a variant of the cross-correlation approach commonly used for relative image registration.  The scaling of the diffraction pattern with wavelength enables this approach to also constrain the absolute centroid, since diffractive scaling expands the PSF about the star's absolute location on the lenslet array. We adopt an iterative approach that ultimately fits all of the data cubes in an imaging sequence together.  We begin by using the central parts of the image to estimate a wavelength-dependent centroid for each cube.  We then transition to using the satellite spots to refine the absolute center position, before performing a global fit to the centroid locations in all data cubes.

Our first step is to compute an approximate centroid for each image and each wavelength using the inner regions of the images.  We adopt the central wavelength channel (typically among the highest signal-to-noise channels) as our reference, and spatially scale all other channels to this reference wavelength.  Each image's diffraction pattern will then have the same size in lenslets but, depending on the center used for the spatial scaling, it will have a translation relative to the image at the reference wavelength.  The value of the translation may be used to derive the centroid of the images.  These are not suitable as final centroids, because the inner regions of the image are strongly affected by the coronagraph and/or saturation (and can be chromatic).  

With an initial guess for the centroid as described above, we next introduce a model for the satellite spot locations.  We assume them to be at a fixed separation and at four angles from the image centroid (a reference angle plus a multiple of 90 degrees).  We further assume that the centroid location itself can be wavelength-dependent, and allow a quadratic shift with wavelength.  This could arise, for example, from uncorrected atmospheric dispersion.  We refine our estimate of the image centroid at each wavelength, including a possible linear shift with wavelength, use the satellite spots.  We optimize for the spot locations that maximize the sum of the interpolated lenslet values at these locations across all wavelength slices. The center of the four spot locations at each wavelength is the image centroid. 

Our next step is to repeat the analysis with the satellite spots, but rather than identifying only the location of peak intensity, we use cross-correlation to match the spot locations in different wavelength channels. This is similar to the cross-correlation approach that we used initially.  However, rather than the entire central region (whose shape can be strongly chromatic because of coronagraphic effects), we limit the cross-correlation to a region within four lenslets of each satellite spot. This stage of our approach uses the actual PSFs of the satellite spots to achieve better precision without having to rely on an analytic model of these spots.  By maximizing the cross-correlation in these limited spatial regions, we derive a new set of centroids at each wavelength, including a possible linear wavelength-dependent drift.  

Our final step combines our centroids with a relative image registration approach.  For relative registration, we derive the best-fit offset between all pairs of images using cross-correlation.  We then perform a least-squares fit to all of the best-fit relative offsets between all pairs of images, obtaining one position per image.  This is only defined up to a constant shift, i.e., these centroids are all relative.  We therefore constrain the mean position to be the same before and after performing this correction.

\subsection{Centroid Methods Comparison}
\begin{figure}
    \centering
    \includegraphics[width=\linewidth]{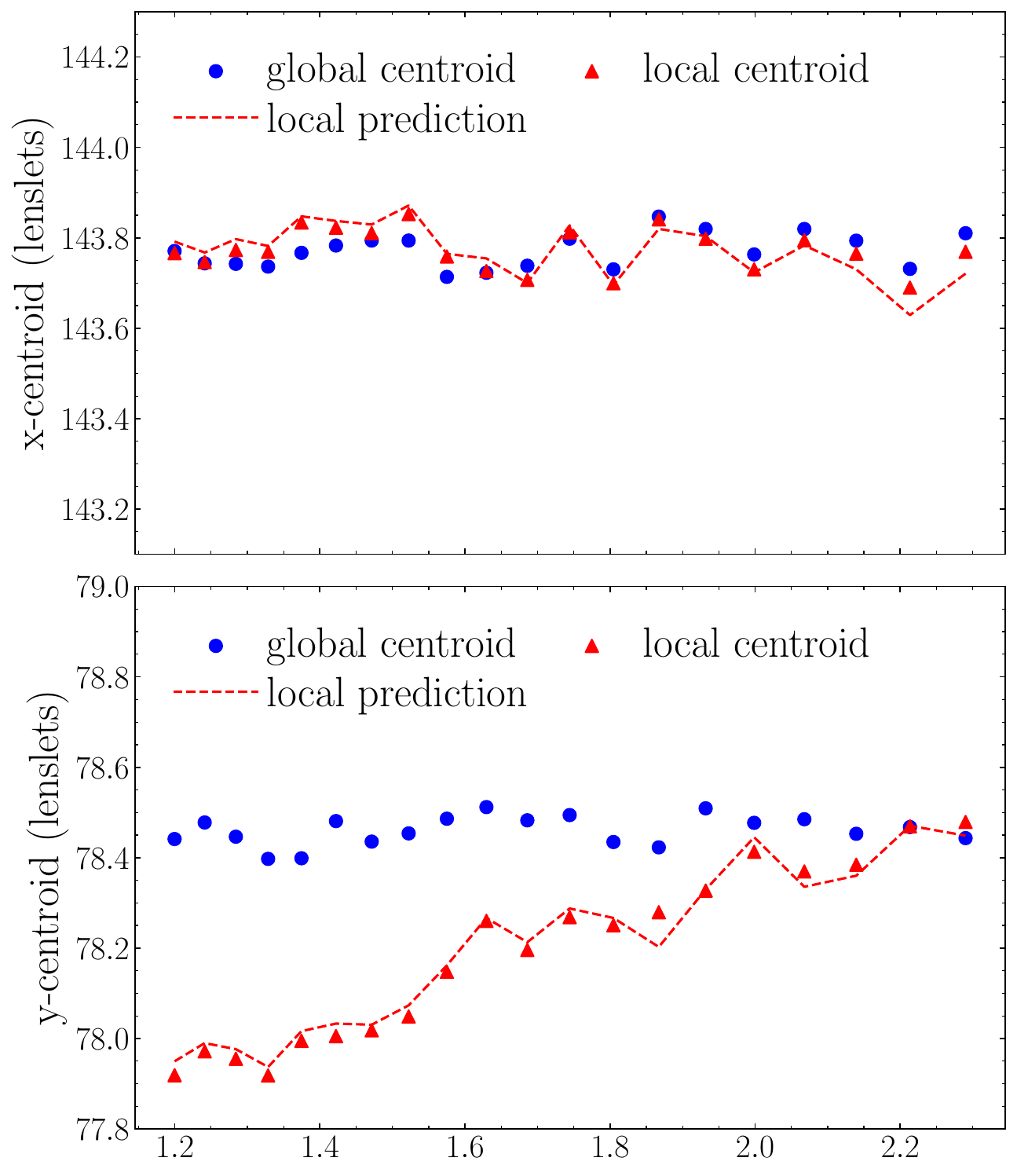}
    \caption{Comparison of global centroid algorithm (blue) vs local centroid algorithm (red). The scatter plot on the top and bottom panels show the best fit \lenslet coordinates of the companion in x and y, respectively. The dashed red line show the expected companion positions based on the blue points, if the centroid measurement differences account for all of the companion position differences.}
    \label{fig:centroid_comparison}
\end{figure}

To test the performance of the global centroid algorithm and compare it to that of the local centroid algorithm, we examine the stability of the position of a bright companion in a \CHARIS dataset under the two centroid algorithms. For this, we use \CHARIS observations of the HD 1160 system taken on UT 2018 December 23rd (PI: Thayne Currie). The dataset has 41 exposures taken over roughly half an hour. Each exposure has an integration time of 31 seconds. The parallactic motion over this period is around 20 degrees. The companion, HD 1160 B, has negligible orbital motion during the sequence of exposures. The companion PSFs in these images are diffraction limited, where the first minimum and the maximum of the Airy Disk are visible even without any post-processing. This allows us to locate the companion position without propagating the data through a post-processing routine, which may introduce additional uncertainties or artifacts that would make it difficult to disentangle the affect of the centroid algorithm alone. We run both centroid algorithms on the dataset. Then we align and derotate the images, and mean-collapse the images in the time dimension. We obtain two combined data cubes this way with high SNRs at all wavelengths, one for each centroid algorithm. We estimate and subtract the background and then fit a 2D Gaussian to the companion at each wavelength of the combine data cubes using MCMC. From this, we obtain best fit positions and widths of the companion PSF at all wavelengths under each centroid algorithm. In principle, the better centroid algorithm would produce companion positions that are more stable over wavelength. The results are shown in Figure. \ref{fig:centroid_comparison}. We find that the companion position is much more stable under the global centroid algorithm, while it drifts by over half a \lenslet under the local centroid algorithm. This indicates that using global information of all the images to constrain the satellite spots simultaneously is more robust and achieves better alignment. To confirm that the drift present in the local centroid algorithm is in fact due to differences in the measured centroids and not due to unknown errors and biases in the PSF fitting of the companion, we also plot the predicted positions of the companion for the local centroid algorithm by translating the centroid differences found by the two centroid algorithms into the differences in the companion positions. These predictions are shown as the dashed red lines Figure. \ref{fig:centroid_comparison}. The agreement between the prediction and the measurements indicates that the drift present in local centroid algorithm is most likely a result of the differences in the two centroid algorithms, rather than unknown errors or biases.

\section{PSF Modeling and Subtraction} \label{sec: PSF Modeling}
\subsection{\klip and Forward Modeling} \label{subsec: KLIP FM}
Once the images are registered and aligned, the next step is to model the diffracted star light that made it past the coronagraph and subtract it from the images. \pyklip does this using the Karhunen-Lo\`eve Image Projection ({\sl KLIP}) algorithm \citep{Soummer2012, Pueyo2015}. The \klip algorithm constructs principal components of the stellar PSF using reference images. For \CHARIS, these are typically images of the same star taken at other times and wavelengths to employ angular differential imaging (ADI) and spectral differential imaging (SDI), respectively \citep{Marois_2006_ADI, Sparks2002}. Images of other stars can also be used to build the principal components for reference differential imaging (RDI), but that approach is not featured in this work. Subtracting the PSF model built using these principal components creates two types of biases: over-subtraction and self-subtraction \citep{FM_Pueyo_2016}. Over-subtractions are corruptions that do not depend on the astrophysical signal. This happens when the algorithm fits the companion with speckle noise and over-subtracts flux from the companion. Self-subtractions are corruptions that depend on the real companion signal in question, and can often be interpreted as subtracting part of the companion signal from itself. These biases can be estimated and corrected for using forward modeling and fake source injection techniques. The products produced by subtracting the PSF model from the data will be referred to as ``reduced'' data/cubes/images henceforth.

\klip is run on small sectors of the image to optimize the stellar PSF modeling on local patches. These sectors are formed by dividing the image into concentric annuli, and dividing each annulus into several equal-sized sectors. The smaller each optimization sector, the more aggressively the data will be fitted.  
\klip also uses a template selection method to model each target optimization section \citep{Marois_TLOCI_2014, FM_Pueyo_2016, pyKLIP_2015}. The movement parameter specifies the exclusion criterion for this selection. The ADI observing mode takes exposures with the image rotator turned off. As a result, a potential source moves through the field of view corresponding to the parallactic angle. If a hypothetical source at the target optimization section would have moved less than the number of pixels specified by the movement parameter at another exposure, then that exposure is excluded from the templates in order to mitigate the effect of self-subtraction. A smaller movement parameter value allows images with smaller field rotations relative to the target image to be included for PSF-fitting, and is more aggressive. Self-subtraction is still present in the PSF-subtracted target section because the companion still remains in all selected templates, only now offset by an amount larger than the movement parameter in the templates compared to the target section. This produces negative wings visible in all \klip reduced images. This is corrected for with forward modeling. Furthermore, the images are typically high-pass filtered with a Fourier-based implementation in order to remove any smooth large-scale features in the image. This way, the \klip algorithm is mainly working on modeling individual speckles, each of which is the spatial scale of the diffraction limit. 

In addition to the size of the optimization sections and the movement criterion, the number of principal components used to construct the model also influences how aggressively the fit to the stellar PSF is. For \pyklip, we typically refer to the principal components as ``KL modes'' and we use ``KL $x$'' to refer to using the first $x$ principal components for the stellar PSF model. Using more KL modes fits for the stellar model/speckle pattern more aggressively. This results in better speckle suppression and, naively, a better detection limit. However, more KL modes also capture more of the signal of any potential companion. This could lead to worse throughput and introduce more over-fitting, which need to be corrected for \citep{Soummer2012, Pueyo2015}. 
The SNR of the companion are often maximized at an intermediate level of aggressiveness, which have been mostly empirical choices in previous works \citep{FM_Pueyo_2016, Wang_2018_HR8799_GPI}. 
We use the default values in the \pyklip package as our starting point to reduce our datasets in this work.  

Any point source we detect in the images after running \klip is distorted due to self-subtraction and over-fitting. To account for this, we use the \klip Forward Modeling (\klipfm) framework \citep{FM_Pueyo_2016} to approximate the over-fitting caused by \klip. A key input for \klipfm is the instrumental PSF, which we can extract from the four satellite spots in the image (see Section \ref{subsec: global centroid}). This enables us to perform accurate PSF fitting to extract information such as the brightness and spectrum of a point source in the data. We will use this in Section \ref{sec: spectra extraction} to extract the spectrum of a point source in the data. 

\subsection{\wlowrank Approximation} \label{subsec: weighted lowrank}

\klip and related methods based on Principal Component Analysis (PCA) construct a set of basis images.  These basis images minimize the squared residual across all pixels and all images used to construct the basis set.  The set of basis images can be efficiently computed using the singular value decomposition (SVD).  The popularity of PCA is partially due to the simplicity of computing the principal components.  

The squared residuals minimized by PCA equally weight all pixels of all images.  This makes it impossible to mask a bad pixel or a planet's location by, for example, setting that weight to zero.  It also means that pixels with higher counts drive the fit, as the same relative residual translates to a much larger absolute residual.  
A more flexible approach to overcome these shortcomings is to minimize the weighted squared residuals
\begin{equation}
\sum_{{\rm images}\,i} \sum_{{\rm pixels}\,j} W_{i,j} \left( I_{i,j} - I'_{i,j} \right)^2
\label{eq:L2norm}
\end{equation}
where the weights $W_{i,j} \geq 0$ may be different for each pixel in each image.  In Equation \eqref{eq:L2norm}, the $I'_{i,j}$ are linear combinations of the basis images.  For example, if the errors on each pixel were known and Gaussian, the weights would be the inverse variances.  Unfortunately, the optimization problem defined by Equation \eqref{eq:L2norm} may not be solved by SVD, or indeed, by any closed-form approach \citep{Srebro+Jaakkola_2003}.  
 
We use the term \wlowrank for this approach.  For the general weighted low-rank approximation, iterative algorithms are the only options \citep{Srebro+Jaakkola_2003}.  \cite{Bailey_2012} has implemented one such approach, the expectation-maximization algorithm, or \empca.  \cite{Bailey_2012} applied the algorithm to construct a set of basis spectra for quasars; we turn to the problem of basis images for high-contrast datasets.  We adapt the software of \cite{Bailey_2012} for use in the \CHARISpipeline.  

The weighted low-rank approximation makes it straightforward to mask a planet's location in any dataset by setting the weights of the appropriate pixels to zero.  Hot pixels and cosmic rays may be handled the same way, without any need for interpolation.  Different weights due to photon and background noise are also straightforward to include.  

By setting the weights to zero at pixels where a planet appears in any individual image, we may guarantee that none of the planet's flux will contribute to the basis images used for PSF subtraction.  In other words, it is possible to completely eliminate self-subtraction.  This is not possible with \klip, as the pixels containing planet light will be in different positions at different times due to parallactic motion.  Masking them requires an approach that can handle missing data at any position in any part(s) of a dataset.  We can then use these same weights to fit the basis images to the science frames.  This step prevents overfitting to enable an unbiased extraction of the planet's flux or spectrum. An example of this is shown in Section \ref{subsec:epmca_vs_klip}, for an injected synthetic source.

A complete elimination of self-subtraction and oversubtraction requires masking the location of a planet out to several pixels beyond its peak intensity.  This reduces the fidelity of the approximation of the stellar PSF, and results in a tradeoff between removing subtraction artifacts and maximizing PSF suppression.  The mask must be recomputed for each possible planet location, which also makes it computationally costly to apply to a full dataset.  

\wlowrank offers an alternative, complementary data reduction approach to extract spectra with fewer artifacts and smaller required corrections than PCA-based approaches.  In fact, PCA is a special case of \wlowrank (with uniform weights), so there is guaranteed to be some region within the vast parameter space of joint pixel weights for which \wlowrank outperforms PCA. Our implementation of \wlowrank in \pyklip enables users to apply it to any dataset that \pyklip supports, not just exclusive to CHARIS.  Section \ref{subsec:epmca_vs_klip} compares the performance of \wlowrank to that of \klip for a dataset of the planet system around HR~8799. We restrict this comparison to a single case of weights that mask the location of a companion. A full investigation of the possible space of pixel weights is beyond the scope of this paper.

\section{Spectral Extraction} \label{sec: spectra extraction}
Once a point source is detected in a dataset, the process to extract its spectrum is two-fold, but both use the \klipfm features implemented in \pyklip. First, we measure the position of the source in the wavelength-collapsed broadband image using \klipfm and the PSF fitting procedure described in \citet{Wang2016}. This generally enables us to measure the position of the point source well below sub-pixel accuracy, even after accounting for uncertainties due to correlated speckle noise in the image. The second step follows the algorithm described in \citet{Greenbaum_2018}: we fix the position of the point source we want to forward model, and use \klipfm again to fit the brightness of the point source in each individual spectral channel. This procedure accounts for how the planet PSF at a given wavelength distorts itself at all other wavelengths due to SDI. Documentation for how to do this full procedure on \CHARIS data is available at the \pyklip docs.\footnote{\url{https://pyklip.readthedocs.io/en/latest/instruments/CHARIS.html}}

To measure the uncertainties on the extracted spectrum, we inject point sources at the same separation as the real companion but at many other position angles and then measure the spread of the recovered spectra for these injected sources. The uncertainties measured this way are those of the PSF subtraction and spectral extraction. These are then propagated with systematic calibration uncertainties from other aspects such as those of the satellite spot contrasts and those of the observed stellar magnitudes. Furthermore, since \klipfm corrects for the bias of self- and over-subtraction based on a perturbation-based linear expansion \citep{FM_Pueyo_2016}, it may still have significant biases for aggressive reductions for which non-linear effects become more prominent, especially for the brightest targets. Therefore, we also use the injected sources and the recovered spectra to measure the non-linear biases and correct for them.

\section{Spectrophotometric Calibrations} \label{sec: specphot cal}
After extracting the companion contrast spectrum relative to the satellite spot PSF as described in the previous section, we need to carry out spectrophotometric calibrations to convert the contrast spectrum into physical flux density units. 

To do this, we first convert the companion contrasts relative to the satellite spots into contrasts relative to the host star using the satellite spots to host star contrasts:
\begin{equation}
    \frac{F_{\rm target, \lambda}}{F_{\rm star, \lambda}}=\frac{F_{\rm target, \lambda}}{F_{\rm spot, \lambda}} \frac{F_{\rm spot, \lambda}}{F_{\rm star, \lambda}}
\end{equation}
where $F_{\rm object, \lambda}$ denotes the flux density of the object at wavelength $\lambda$. Note that each individual flux density in this equation is not yet known, only the fractions, i.e.~the contrasts, have been calculated.
\cite{Currie_2020SPIE} used \SCExAO's internal source over a narrow bandpass centered on $1.55 \mu$m to measure the contrasts of the satellite spots relative to the internal source. They measured a contrast of $\frac{F_{\rm spot, \lambda}}{F_{\rm star, \lambda}}=2.72\times 10^{-3} \pm 1.3 \times 10^{-4}$ at $1.55 \mu m$ at a $25$\,nm modulation amplitude of the diffractive grid on the \SCExAO deformable mirror. The intensity of the satellite spots, and hence the contrast, should scale as $\propto A^2\lambda^{-2}$, where $A$ is the modulation amplitude of the diffractive grid, and $\lambda$ is the wavelength. \cite{Currie_2020SPIE} found excellent agreement with the $\propto \lambda^{-2}$ scaling in the narrowband of $1.1 \mu m -1.8\mu m$ (agree to within $\sim 1\%$ or less), as well as excellent agreement with the $\propto A^2$ scaling. However, they note a difference on the order of $\sim 8\%$ between separate tests which could plausibly be due to slight changes in the placement of \SCExAO's focal-plane masks between the times of the two tests. In addition, \cite{Wang_2022} also found correlated fluctuations in the satellite spot fluxes over the course of an observing sequence, as well as a $3\%$ difference in the average satellite spot contrast compared to \cite{Currie_2020SPIE}. As a consequence, high precision spectrophotometric studies are not yet feasible with \SCExAO/\CHARIS. To minimize the risk of such discrepancies, we recommend the user look into the satellite spot contrast measured on a date closest to the observation date of the dataset the user is analyzing. The satellite spot contrast measured by \cite{Currie_2020SPIE} quoted above is recorded as a python class constant in the \CHARIS module in the \pyklip package. It can be trivially assigned to other values if the user finds better suited calibration tests for the satellite spot contrasts. Details of accessing and handling these constants can be found in the \CHARIS documentation online.

We then convert the companion's contrast spectrum relative to the host star to a companion spectrum in physical flux density units using information about the host star. Given the star's spectral type, and optionally its temperature and surface gravity, we obtain a stellar model spectrum of the host star. We scale this stellar model to physical flux density units using the observed magnitudes of the star and properly re-sample them at the \CHARIS wavelength bins to obtain $F_{\rm star, \lambda}$. The companion spectrum in flux density units is then
\begin{equation}
    F_{\rm target, \lambda} = F_{\rm star, \lambda}\frac{F_{\rm target, \lambda}}{F_{\rm star,\lambda}}
\end{equation}

For the stellar model, the Pickles library \citep{Pickles_1998} is widely used. However, for some spectral types, the library of Pickles lacks direct measurements in the near-IR and extrapolates from shorter wavelengths that could lead to miscalibrated companion spectrum \citep{Currie_2018_kappa}. Therefore, we follow the same choice made in \citealt{Currie_2018_kappa} and use the Castelli and Kurucz stellar models \citep{Castelli_Kurucz_2004} as the default option for our pipeline, implemented via the package {\tt pysynphot} \citep{pysynphot_2013}. The Pickles library remains as a back-up option for the calibration as well.

\section{Full Reduction Demonstrations} \label{sec: full reductions}
In this section, we run the full pipeline reduction on various injected spectra as well as on \CHARIS data with known real companions to demonstrate the performance of the pipeline. As mentioned in section \ref{sec: pipeline overview}, CHARIS has broadband (J, H, and Ks) data and narrowband (J, H, or Ks) data with slightly higher resolution. The reduction process is identical and we will only demonstrate the results for broadband data in this paper. We also limit the scope of this paper to point sources. 

\subsection{Contrast Curve} 
\label{subsec: contrast curve}
In this section, we show the calibrated contrast curves for two \CHARIS observing sequences taken on Sept 1st 2018 and Aug 31st 2020. We also demonstrate the pipeline's performance on known, injected spectra in this section.

We first measure three broadband contrast curves to see how the detection limit depends on the angular separation and integration time for \CHARIS. We use two \CHARIS datasets to generate these contrast curves: one dataset of HR~8799 (PI Jason Wang) taken on Sept 1st 2018 that covers a parallactic angle rotation of $\sim 165\degree$ \citep{Wang_2022}; one dataset of HD~33632 (PI Thayne Currie) taken on Aug 31st 2020 that covers a parallactic angle rotation of $\sim 18\degree$. The apparent magnitudes of the host stars and observing conditions are similar for both systems. As a result, the AO corrections are good for both datasets and are of similar quality visually. We also take a subset of the HR~8799 dataset that covers about the same amount of field rotation as the HD~33632 dataset and measure the contrast curve on this subset to examine the consistency of the detection limit for different observation dates with different conditions. 

\begin{figure}
    \centering
    \includegraphics[width=\linewidth]{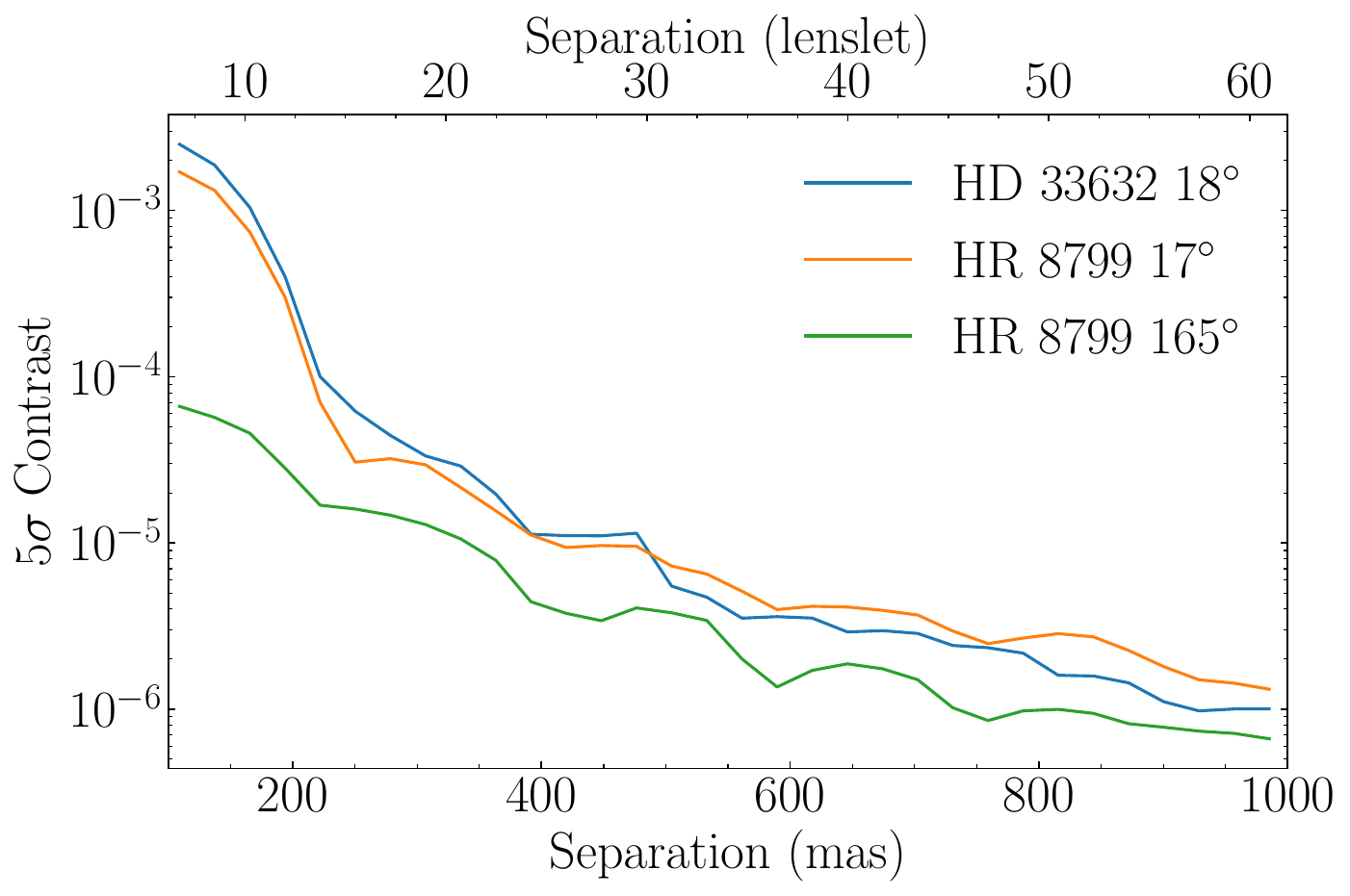}
    \caption{Calibrated contrast curve showing the theoretical $5\sigma$ detection limit as a function of separation for a typical CHARIS reduction.}
    \label{fig: calibrated contrast curve}
\end{figure}

To measure the contrast curve, we use the reduced, wavelength-collapsed image along with tools implemented in the \pyklip package. First, \pyklip assumes azimuthally symmetric noise and compute the $5\sigma$ noise level at each separation, smoothing out high frequency noise using Gaussian cross correlation and account for small number statistics. This gives us an estimate of the detection limit at each separation, without accounting for the algorithm throughput. Based on this estimate, we inject fake sources a few times brighter than the detection limit, run \klip reductions and measure the fluxes of these sources to estimate algorithm throughput. We inject fake sources at three different separations: 20 pixels, 40 pixels and 60 pixels. At each separation, we inject the same source at four different PAs. We average the throughput results over PA to obtain throughput estimates at the three different separations. Then, we calibrate the previous contrast curve by correcting for the throughput as a function of separation. The calibrated contrast curves are shown in Figure \ref{fig: calibrated contrast curve}, which provides detection limit estimates as a function of angular separation within the field of view of \CHARIS. We see that two datasets with similar observing conditions produce similar contrast curves, while a longer observing sequence with more field rotation improves the detection limit as expected.

\subsection{Performance on Injected Spectra}
\label{subsec: injected performance}
We now demonstrate the pipeline's performance on extracting companion spectra using injected sources. For the injected synthetic sources, we use substellar template spectra of three different spectral types---M7, L7 and T1---to represent the typical range of brown dwarfs being observed \citep{Burgasser_2004_Dwarf_Templates, Burgasser_2010_Dwarf_Templates}. In addition, we also test the pipeline on a flat spectrum to represent sources with few spectral features such as white dwarfs. We inject fake spectra with three different brightnesses: similar brightness to that of the speckles, $\sim 5$ times fainter than the speckles, and $\sim 10$ times fainter than the speckles. We choose these brightnesses to cover a full range of observations from bright companions to fainter ones just above the detection limit. We show that the pipeline can successfully retrieve spectra in all these scenarios with the appropriate selection of reduction parameters. We inject all spectra at a separation of 45 pixels or $\approx$725 \hbox{mas}.

\begin{figure}
    \centering
    \includegraphics[width=\linewidth]{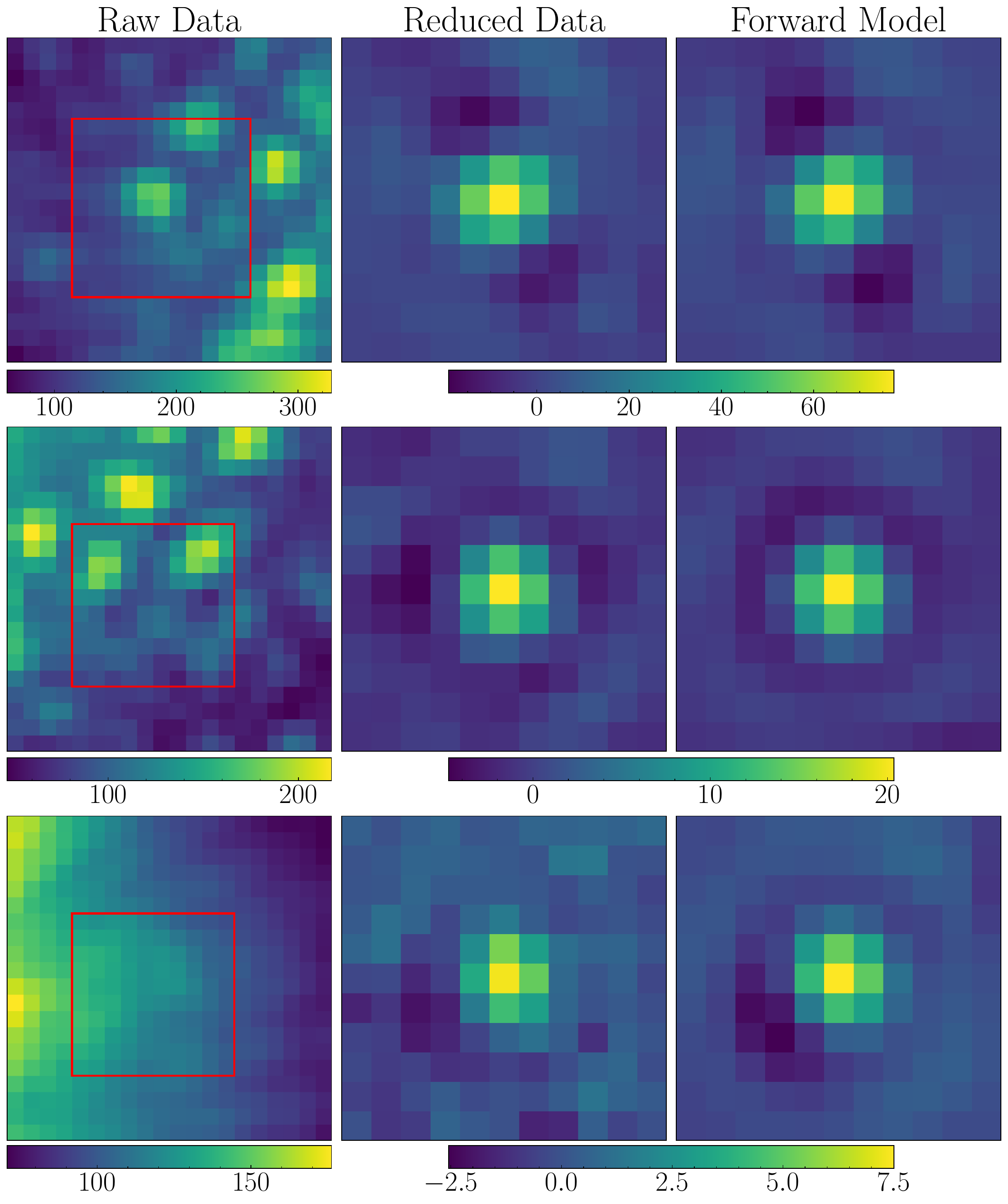}
    \caption{Examples of the raw image, reduced image, and the forward model for the full reduction of an injected source with similar brightness (top panel), $\sim$ 5 times fainter brightness (middle panel), and $\sim $ 10 times fainter brightness (bottom panel) than the speckles. The raw images are zoomed out to include the speckle pattern more clearly. The middle and right columns are zoomed in on the red squares in the left column.}
    \label{fig: all injected images}
\end{figure}

Figure \ref{fig: all injected images} shows the images at one wavelength slice of the injected L7 template as an example. From left to right columns are images before reduction, the PSF-subtracted image, and the forward model of the region around the injected template. The effect of over-subtraction is clearly visible as the negative wings around the source; the forward model is shown to account for this effect. From top to bottom panels are the different brightnesses: similar brightness to that of the speckles, $\sim 5$ times fainter than the speckles, and $\sim 10$ times fainter than the speckles, respectively. 

\begin{figure}[p]
    \centering
    \includegraphics[width=0.95\linewidth]{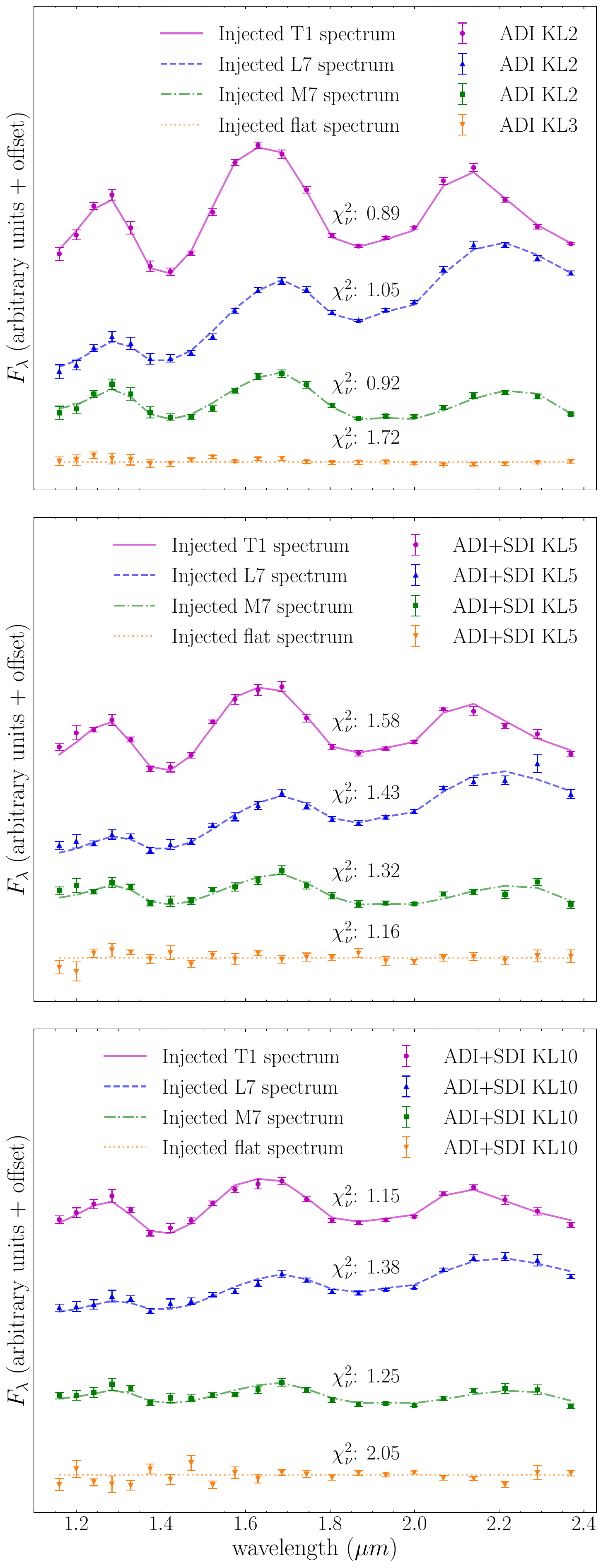}
    \caption{The spectral extraction results for four types of substellar templates injected at three different brightnesses: similar brightness (top panel), $\sim$ 5 times fainter brightness (middle panel), and $\sim$ 10 times fainter brightness (bottom panel) than the speckles. The data points with error bars are the extracted spectra. The y axes are in flux density units, but are offset by an arbitrary factor for a clear visual.}
    \label{fig: all_inejcted_spectral_extraction}
\end{figure}

Figure \ref{fig: all_inejcted_spectral_extraction} shows the four types of extracted spectra: three substellar templates and a flat spectrum, calibrated with both forward modeling and bias correction described in sections \ref{subsec: KLIP FM} and \ref{sec: spectra extraction}, respectively. The top to bottom panels are again the three different brightnesses. For the top panel, we use a ADI only reduction and a small KL mode number because the injected source is very bright. The extracted spectra all agree well with the true injected spectra. The middle panel shows the same figure as the top but now for templates that are $\sim$ 5 times fainter than the speckles. For this reduction, we use a slightly more aggressive reduction by including SDI along with ADI in the PSF modeling. The accuracy of the extracted spectra compared to the injection is visibly a bit worse than the brighter case, but the error estimates also increased correspondingly. There are still good agreements in terms of reduced $\chi^2$ for this set of extraction. The bottom panel shows the same figure for templates that are $\sim 10$ times fainter than the speckles. This is a few times brighter than the measured $5\sigma$ detection limit. A slightly more aggressive set of reduction parameters is used: 10 KL modes and ADI$+$SDI. The extracted spectrum still has good agreement with the injection, though the recovered flat spectrum now has more significant deviations from the true spectrum, indicating that a featureless spectrum near the detection limit is difficult to recover. 

Overall, these tests on injected spectra show that the pipeline can reliably recover the spectra for bright (lower contrast) companions as well as faint (higher contrast) ones that are near the detection limit.

\subsection{Performance on \CHARIS Data} \label{subsec: full reduction real data}
In this section, we use the pipeline to reduce two real \CHARIS observations of two different systems with previously characterized companions, HR~8799 and HD~33632, and compare our calibrated spectra with the published results to show that the pipeline successfully recovers spectra that agree with the literature for these well-studied sources.

\subsubsection{HD 33632}
\begin{figure}
    \centering
    \includegraphics[width=\linewidth]{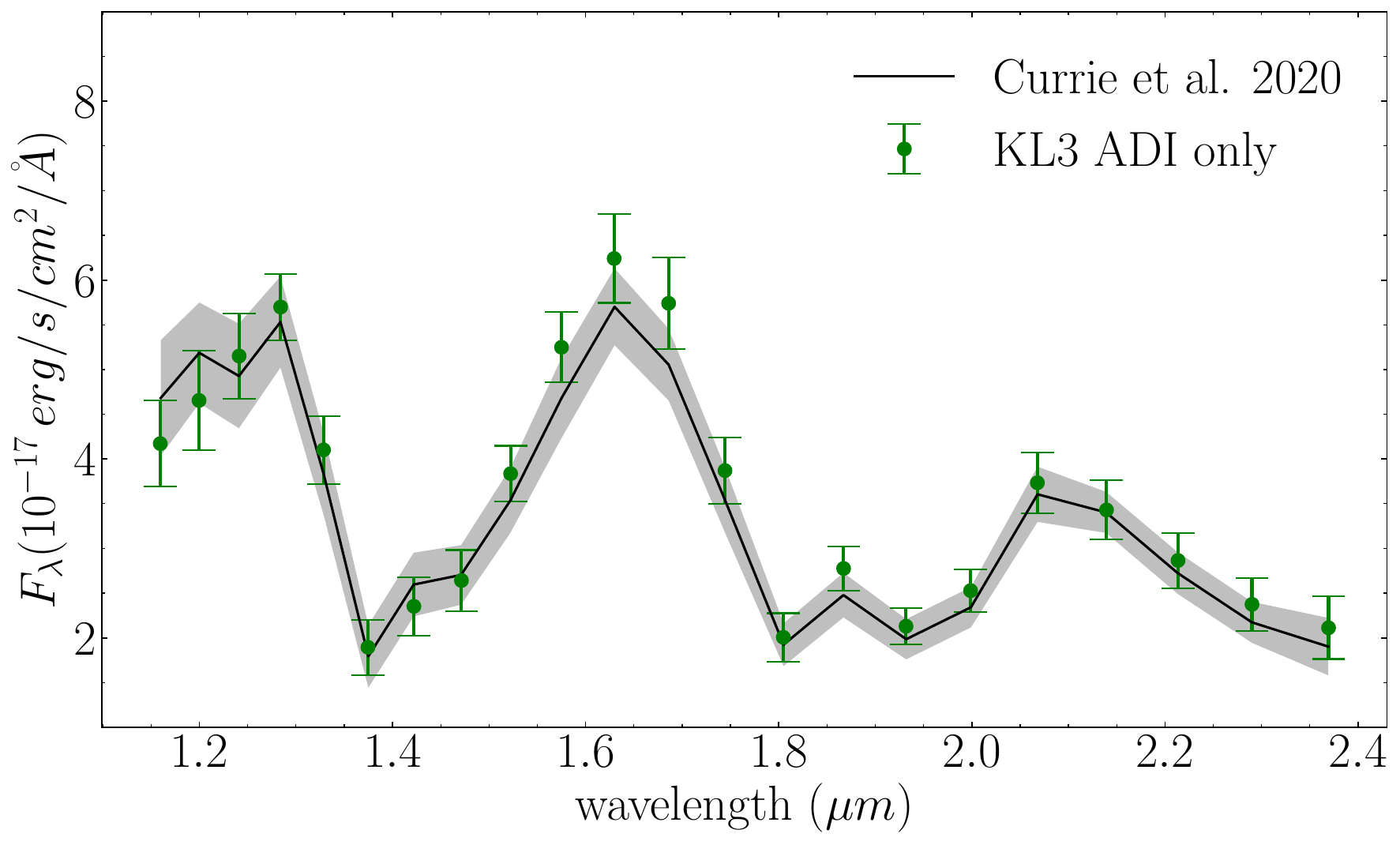}
    \caption{\klipfm extracted spectra of HD~33632 Ab at three similar KL modes compared to the calibrated spectra published in \citealt{Currie_2020_HD33632}.}
    \label{fig: HD33632_calibrated}
\end{figure}

We first show the pipeline extracted spectrum for a substellar companion to a nearby Sun-like star, HD~33632 Aa. The companion, HD~33632 Ab, was detected in direct imaging by \cite{Currie_2020_HD33632} using \SCExAO/\CHARIS. The data cubes were extracted using the \CHARIS DRP \citep{Brandt_2017_CHARISPipeline} and then PSF-subtracted and calibrated using the IDL-based \CHARIS DPP, which utilized the A-LOCI algorithm in this case \citep{Currie_2012_A-LOCI}. We use the same \CHARIS dataset \cite{Currie_2020_HD33632} used in their discovery paper, but now we use our pipeline to perform the analysis. The contrast for this companion to its host star in the \CHARIS broadband (JHK) is around $5 \times 10^{-5}$ in a wavelength collapsed image, about an order of magnitude above the detection limit according to the contrast curve shown in Figure \ref{fig: calibrated contrast curve}. Therefore, a non-aggressive reduction is ideal for this companion in order to obtain a good SNR with as little over-fitting as possible. We choose a few small KL mode numbers and run \klip using ADI only. We divide every image into 9 annuli and 4 sections for optimization. We then fit for the companion location using the forward model. Finally, we extract and calibrate the spectrum as described in section \ref{sec: spectra extraction} and \ref{sec: specphot cal}. In Figure \ref{fig: HD33632_calibrated}, we show the \pyklip extracted spectra compared to the results published in \citealt{Currie_2020_HD33632}. The error bars for our analysis are calculated using the fake source injection method described in section \ref{sec: spectra extraction}. The \CHARISpipeline is able to reproduce the published spectrum with good agreement (well within $1\sigma$ across the entire \CHARIS broadband). This agreement is particularly reassuring because the two pipelines use different image registration and PSF-subtraction algorithms, as well as independently implemented spectrophotometric calibrations. This cross-check adds confidence to analyses that depend on the extracted shape of the companion spectrum.

\subsubsection{HR 8799}
The second dataset we use to demonstrate the pipeline performance is a \CHARIS broadband dataset of the HR~8799 system \citep{Wang_2022}. HR~8799 is a $1.5$ \Msun ``F0'' main sequence star \citep{Gray_1999, Gray_2003, Baines_2012} with four imaged planets \citep{Marois_2008, Marois_2010}. This is one of the most famous and well studied directly imaged planetary systems. Three of its four imaged companions, HR~8799 c, d, and e, are within the \CHARIS field of view. We extract the spectra for these three planets using the same procedures as for HD~33632 and compare the extracted spectra with the Gemini Planet Imager ({\sl GPI}) spectra published in \citep{Greenbaum_2018}. The \CHARIS broadband contrasts of planet c, d and e to the host star are roughly $1.7 \times 10^{-5}$, $1.7 \times 10^{-5}$ and $1.2 \times 10^{-5}$ respectively. They are more extreme than the contrast for the HD~33632 system. For the case of HR~8799 e, in particular, the contrast is approaching the detection limit of \CHARIS due to the small separation ($\sim$25 pixels). Therefore, we use both ADI and SDI to model the PSF, and also increase the KLmodes used for each of these planets accordingly.

\begin{figure}
    \centering
    \includegraphics[width=\linewidth]{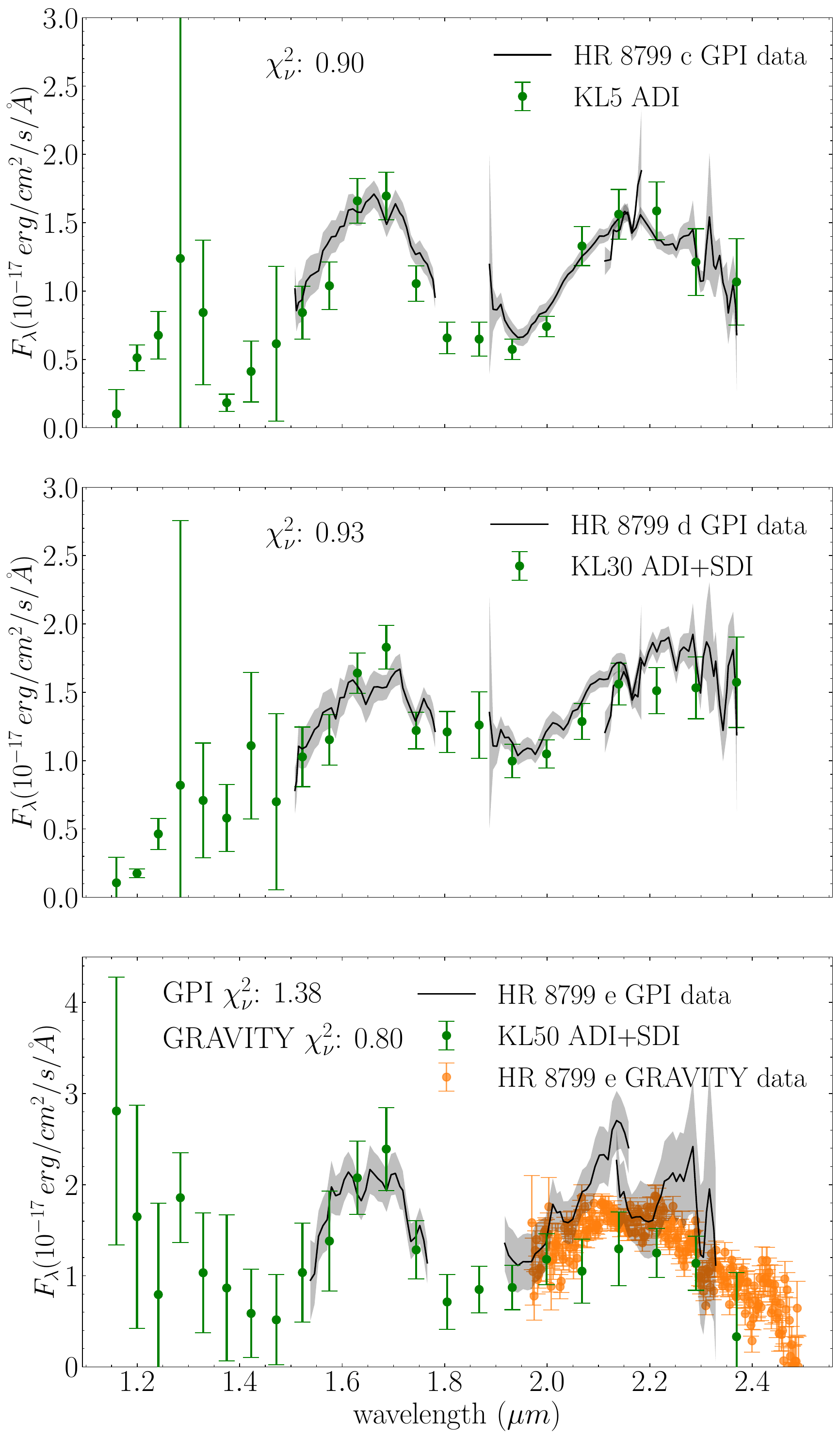}
    \caption{\klipfm extracted spectra of the three planets HR~8799 c, d and e, compared to the calibrated \GPI spectra for these planets published in \citealt{Greenbaum_2018}.}
    \label{fig: HR8799_calibrated}
\end{figure}

Figure \ref{fig: HR8799_calibrated} shows the comparison of our extracted spectra for HR~8799 c, d and e with the \GPI spectra. The \GPI spectra were also extracted using \pyklip, but with separate image registration and calibrations unique to \GPI \citep{Greenbaum_2018}. The \GPI spectra cover the H and K bands, with two separate filters K1 and K2 in K band \cite{GPI}. There is a small overlap between K1 and K2 which can be seen in the Figure. For the outer two planets, HR~8799 c and HR~8799 d, our spectra agree with the \GPI spectra well with reduced $\chi^{2}$ close to 1. For the inner most planet, HR~8799 e, the fluxes in H band again show good agreement between \GPI's results and ours. The agreement is worse in K band, particularly for K1 ($\sim 1.9 \mu m-2.2 \mu m$). We suspect this is an issue with the \GPI instrument, as there is also a discrepancy within \GPI's own spectrum, where there is a gap between K1 and K2 for HR~8799 e. This discrepancy is not present for the outer two planets. This could be due to large residuals of the forward model fitting for HR~8799 e, especially in K1 band \cite{Greenbaum_2018}. Therefore, to compare to an additional reference, we also included \GRAVITY's spectrum for HR~8799 e \citep{GRAVITY_collaboration_2019}, and we find that \CHARIS's spectrum is consistent with \GRAVITY's spectrum with a reduced $\chi^{2}$ of 0.8. 

To summarize, we find our spectra for the three HR~8799 planets within the field of view of \CHARIS to be consistent with those extracted from \GPI's data (with the exception of HR~8799 e in K1 band, but in agreement with \GRAVITY's spectrum). This serves as a consistency check for the image registration and calibration steps implemented in the \CHARISpipeline, which share the same PSF subtraction and forward modeling implementation as \GPI's \pyklip pipeline.

\subsection{\empca vs \klip} \label{subsec:epmca_vs_klip}

In this section we compare the spectra of injected sources extracted using \klip and \empca. \empca currently only implemented ADI and does not have SDI or forward modeling capabilities, it is expected to be strictly inferior to \klipfm in most situations, as our results will show. However, while implementing forward modeling for \empca is a future effort that is beyond the goal of this paper. We show some of its unique advantages when compared with \klip using ADI-only reductions without forward modeling, and highlight its potential to motivate future effort to develop forward modeling and optimized pixel weights.

As mentioned in Section \ref{subsec: weighted lowrank}, \empca is an implementation of the \wlowrank Approximation, which is a more general form of PCA that allows individual pixel weights when minimizing the residuals. Therefore, there is guaranteed to be some region within the parameter space of pixel weights that outperforms PCA, which has fixed uniform weights. However, systematically explore and optimize the pixel weights is a task beyond the scope of this paper. We explore several simple choices of pixel weights, $W_{i,j}$, from Equation \ref{eq:L2norm} and compare them to a KLIP reduction as a demonstration of how the weights can affect the reduction in terms of SNR and spectral fidelity. 

We inject a fake source 5 times fainter than the background speckle pattern the same way we did in section \ref{subsec: injected performance}. We run ADI reductions with \empca and \klip, using the same numbers of annuli and subsections. This means the images are divided in to the same optimization sections for both algorithms. But the two algorithms utilize the movement parameters differently. For \empca, instead of using the movement parameter for template selection as described in section \ref{subsec: KLIP FM}, an aperture at the guess location (or known location from a positive detection) of the companion with a radius corresponding to the movement parameter is masked out (i.e. the pixel weights, $W_{i, j}$, for the aperture are set to zeros), and this mask tracks the parallactic angle such that the same physical position in terms of separation and position angle from the central star is masked out for every exposure. Therefore, the companion will be masked out in all exposures in \empca, whereas \klip only ensures the companion in the templates are not overlapping with the companion in the target exposure. This means that \empca could have almost no self-subtraction given a sufficiently large mask over the companion, while \klip would still have negative wings produced by subtracting the offset companions in the templates.

\begin{figure}
    \centering
    \includegraphics[width=\linewidth]{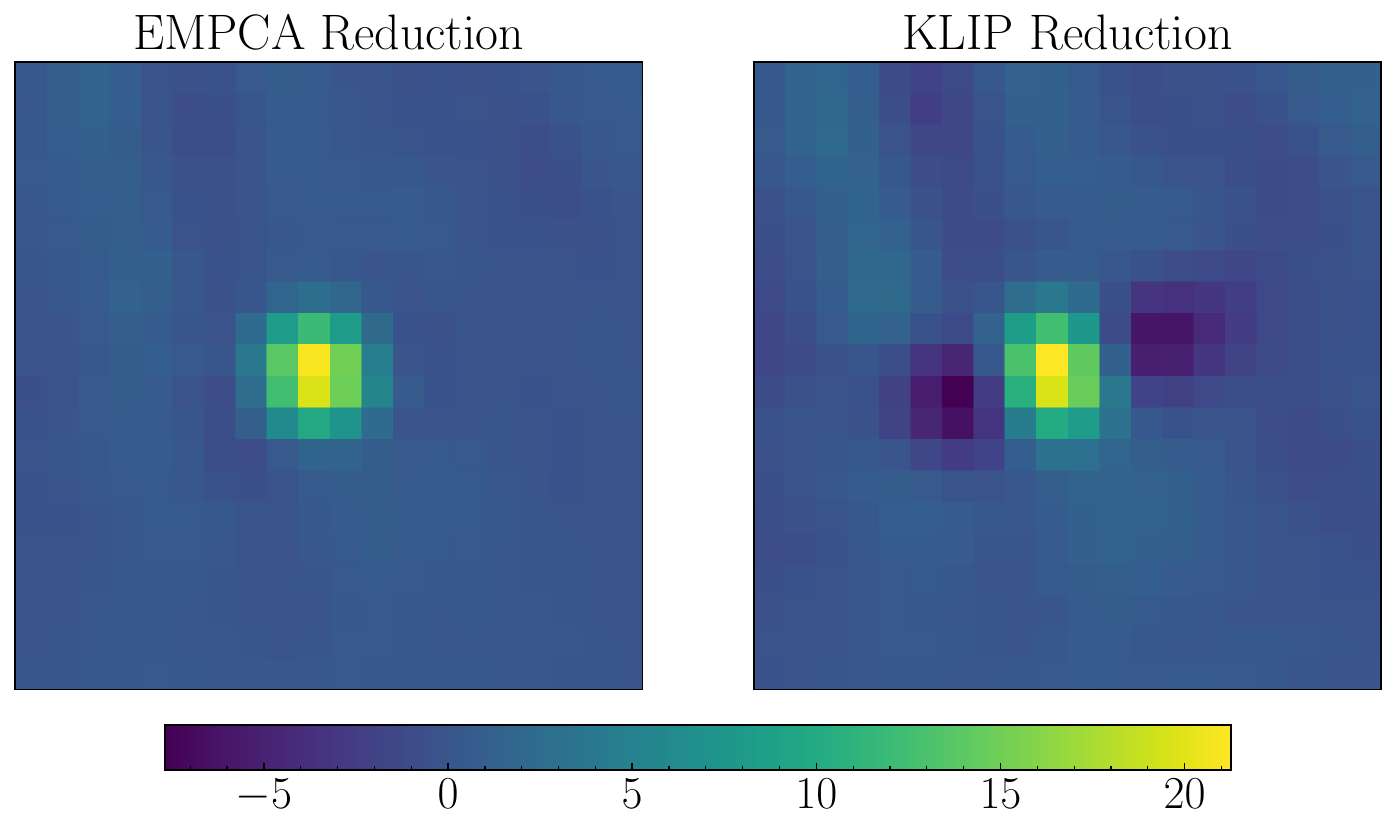}
    \caption{PSF-subtracted broadband images (wavelength averaged) of the injected point source that is roughly $5$ times fainter than the speckles for \empca and \klip reductions.}
    \label{fig: empca vs klip image}
\end{figure}

\begin{figure}
    \centering
    \includegraphics[width=\linewidth]{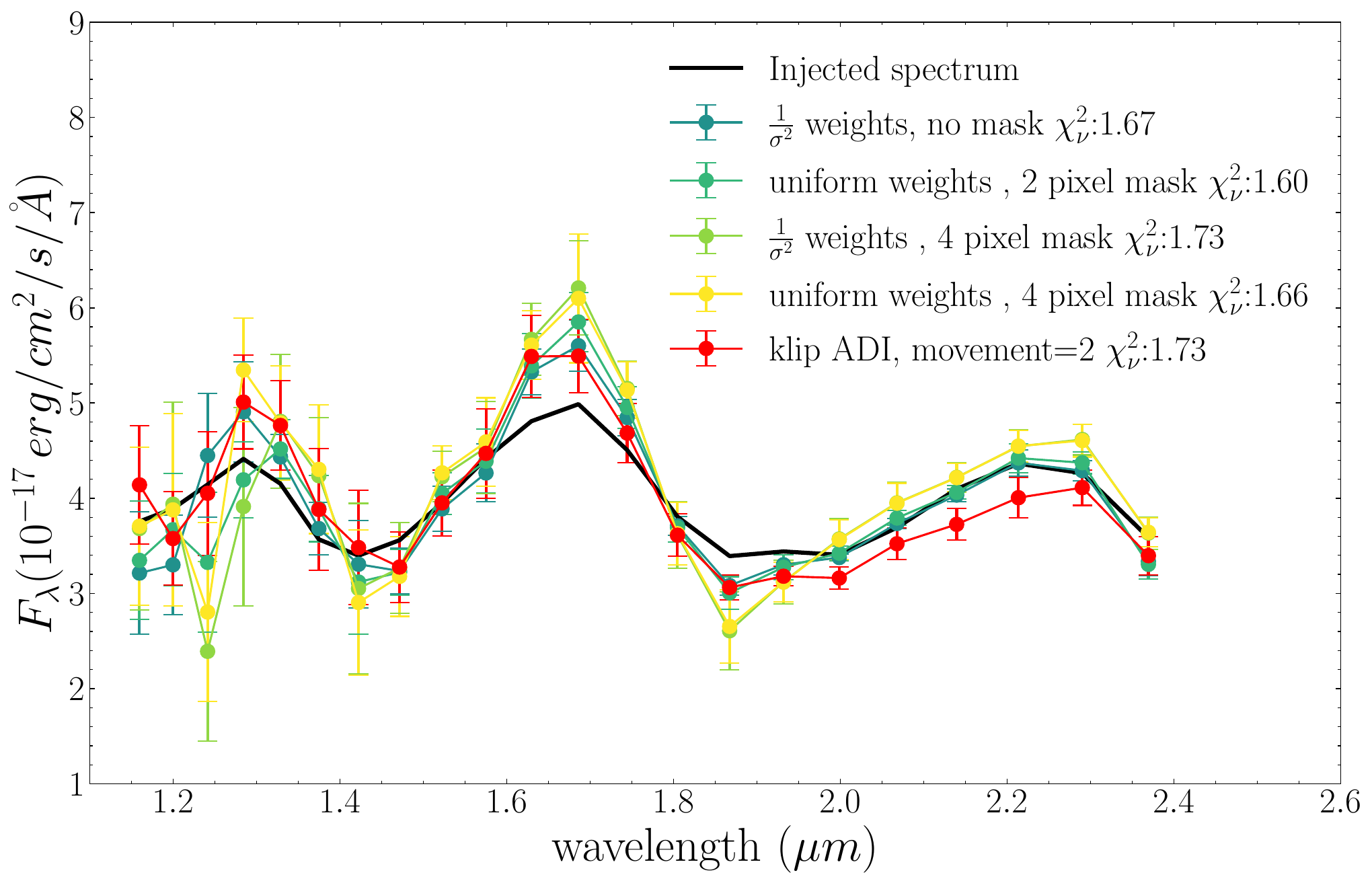}
    \caption{Comparison of \empca and \klip extracted spectra (ADI only, no forward modeling) of the injected point source that is roughly $5$ times fainter than the speckles. The extracted spectra are calibrated using a sample of fake source injections as described in Section \ref{sec: spectra extraction}.}
    \label{fig: empca vs klip spectrum}
\end{figure}

For \empca reductions, we explore a number of different pixel weighting schemes. On a full image scale, we only test two simple choices of weights: 1. Uniform weights , which is equivalent to the weighting in \klip PCA; 2. Inverse variance weights. On top of this, we explore the addition of a local companion mask, a circular region centered at the given separation and PA for which the pixel weights are set to 0. The extracted spectra have been calibrated for biases using the fake source injection method described in Section \ref{sec: spectra extraction}. Figure \ref{fig: empca vs klip image} shows an example of PSF-subtracted broadband images using \empca with a 2-pixel radius mask compared to \klip with 2-pixel movement. And it shows how the mask in \empca is able to effectively eliminate self-subtraction. But after applying corrections for over-subtraction and self-subtraction, measured from the injections of many synthetic sources, the accuracy of \empca extracted spectra in terms of reduced $\chi^{2}$ is not too different from that of \klip, though still mostly comparable or slightly better, as shown in Figure \ref{fig: empca vs klip spectrum}. Because we are not using SDI or forward modeling in order to compare \empca in its current state with \klip on level grounds, we expect the accuracy of the extracted spectra to be worse than that of \klipfm. Spatially varying speckles could introduce bias in this case that are also correlated over wavelength. For our random choice of injection position, it resulted in some excess in H band. However, we only aim to demonstrate that there is room to further optimize the PSF models if we generalize the PCA algorithm to allow free pixel weights using the \empca implementation. 

\begin{figure}
    \centering
    \includegraphics[width=\linewidth]{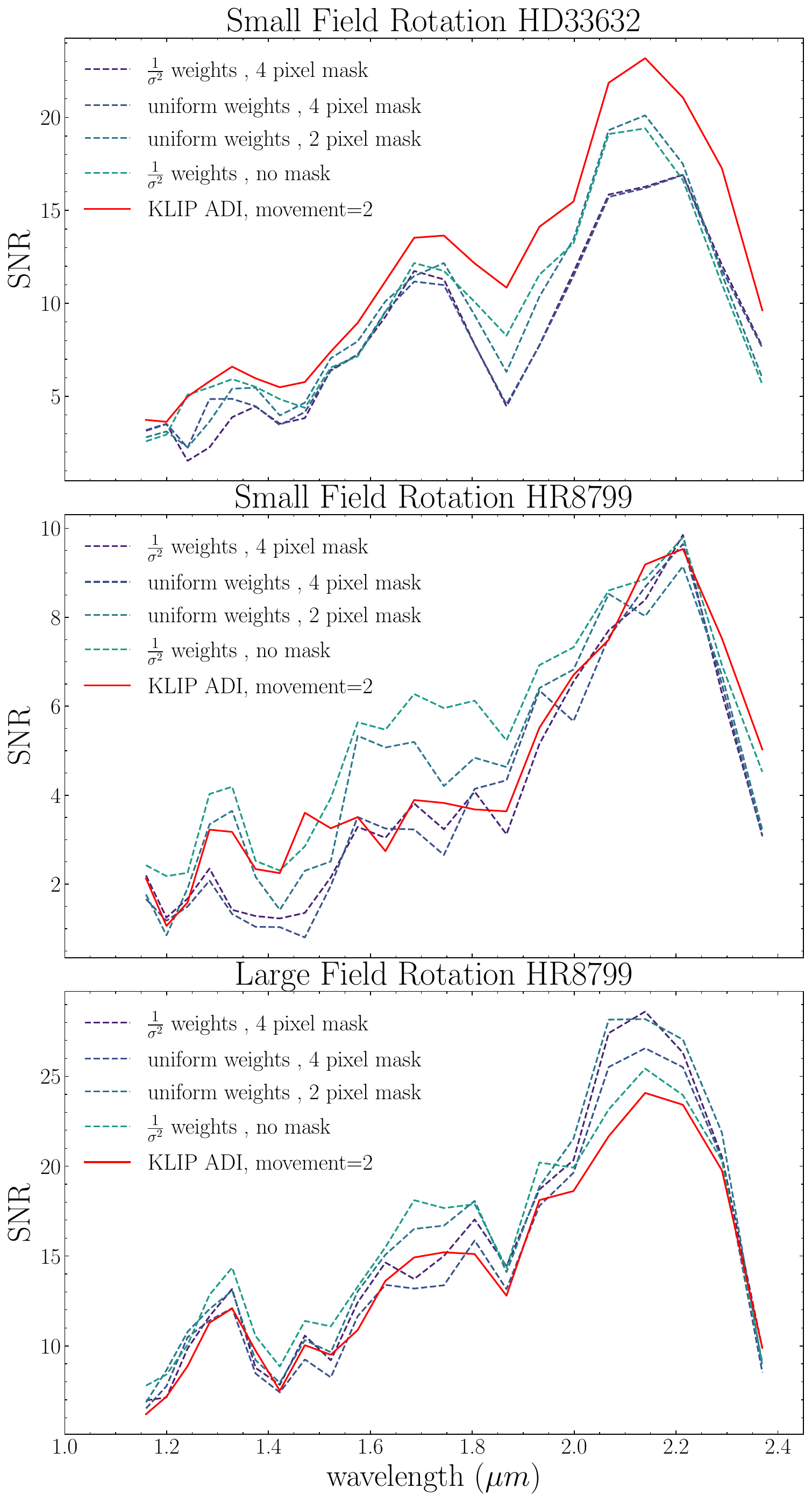}
    \caption{Comparisons of SNRs of an injected point source for \empca vs \klip reductions in three datasets. The top panel shows the results for the source injected into the HD~33632 dataset, which covers a field rotation of $\sim 18 \degree$ from first to last exposure. The middle panel shows the same comparison for the ``slow rotating'' HR~8799 dataset that also covers the same field rotation. The bottom panel is for the ``fast-rotating'' HR~8799 dataset that covers a field rotation of $\sim 160\degree$.}
    \label{fig: empca vs klip SNR}
\end{figure}

A drawback of masking out the companion is that it would decrease speckle suppression within that mask. And if this penalty outweighs the reward of having less self-subtraction, then the SNR of the companion could be worse. We test this on an injected source on the two real datasets described in Section \ref{subsec: full reduction real data}. The the HD~33632 dataset was taken on UTC 2020 Aug 31st (PI: Thayne Currie), with an exposure time of 31 seconds for each exposure. The whole sequence covers a parallactic field rotation of $\sim 18\degree$. The much longer sequence on HR~8799 was taken on UTC 2018 Sept 1st (PI: Jason Wang), with an exposure time of 20.7 seconds, and covers a field rotation of $\sim 160\degree$. These two datasets have different configurations, observing conditions, and total rotation angles, all of which could influence the performances of the two algorithms. Among these factors, the rotation angle is something we can separate from the other factors and control artificially in this analysis. To do this, for the HR8799 dataset, we modified the parallactic angles of the exposures and injected the point source at positions that simulated a slow and a fast rotating sequence, resulting in two datasets that covered different amounts of field rotation, with one dataset covering $\sim 160\degree$ (the original location with un-modified parallactic angles), and the other covering a parallactic rotation of $\sim 18\degree$. As a result, we have a total three datasets that we will reduce using the two algorithms for comparison. To measure the SNR for \klip, we fit for the peak flux of the companion, then we use an annulus around the central star with a radius same as the companion separation to estimate the noise. For \empca, we estimate the noise by masking out many empty regions (no companion) in this annulus and estimating the noise in the reduced data using only these masked regions. This ensures that we are treating the signal region and the noise region the same way so that the SNR is measured from a fair estimate of noise. Figure \ref{fig: empca vs klip SNR} shows the comparison of the SNRs on a injected source that is 5 times fainter than the speckles for these three datasets. The top two panels show that when the field rotations are small, depending on various other factors, \empca could produce similar levels of SNR as \klip, or it may pay more penalty for the increased noise than the reward of a less distorted signal, at least for the weights that we explored. The bottom two panels show that, when the conditions are the same, for a longer observing sequence with a larger field rotation, the reward from having less distortion in \empca outweighs the increased noise in the masked region, and yields a better SNR.

Forward modeling for \klip accounts for over-subtraction and self-subtraction effects well, and produces better results than \empca, evident by comparing Figure \ref{fig: empca vs klip spectrum} with the middle panel of Figure \ref{fig: all_inejcted_spectral_extraction} in section \ref{sec: full reductions}. Nevertheless, it is encouraging to see the advantages of \empca compared to \klip without forward modeling, especially because we have only explored the simplest of the pixel weight schemes and there is a lot of room for in-depth analysis on more complex the pixel weights that optimizes the reduction. This also provides motivation to implement both SDI forward modeling for \empca as well in the future. Even at its current state, \empca can still provide benefits over \klip (with or without forward modeling) in certain situations. One of which is when there are exposures with bad pixels or small artifacts. \klip has to discard these exposures as it cannot apply individual weights to pixels, while \empca can simply mask out these bad pixels by setting their weights to zero.

\section{Summary and Future additions} \label{sec: Summary}
We have presented in this work the \CHARISpipeline under the framework of the open source python library \pyklip, which performs PSF modeling and subtraction based on the \klip algorithm, and extracts the astrometry and spectrum of a companion using forward modeling. We have carried out calibrations for the CHARIS plate scale and PA zero point. We conclude that the plate scale is consistent with the initially designed nominal value and is also stable over time. However, the PA zero point varies significantly from observation to observation, and should be re-calibrated using calibration data taken close to the science data observation time, with minimal instrument set up changes. We have also fully integrated CHARIS compatible spectrophotometric calibrations for extracting planet/brown dwarf spectra. We introduced a new global centroiding algorithm for better image registration and alignment of CHARIS data. We also implemented the \empca algorithm that could serve as a potential alternative to \klip in the event of a significant fraction of bad pixels in the data. We have demonstrated that the pipeline can reproduce the spectra of several real companions published in the literature that are extracted with other pipelines and/or from other instruments' data, as well as recover the spectra of injected sources. We limit the scope of this paper to point source reductions with ADI and SDI, and we hope this Python-based pipeline within the framework of the widely used package \pyklip would serve as a complimentary alternative to the IDL-based \CHARIS DPP \citep{Currie_2020SPIE} and increases CHARIS’s user base to produce more science results.

Moving forward, there are several functionalities on the horizon that would be useful to be added to the \CHARISpipeline in the future. These include the capability to perform polarimetric differential imaging (PDI) for observations of disks, forward modeling and SDI capabilities specifically derived and implemented for \empca, and an in-depth exploration of the vast parameter space of pixel weights for \empca to optimize \empca reductions.

\section*{Acknowledgements}
This research is based in part on data collected at the Subaru Telescope, which is operated by the National Astronomical Observatory of Japan. We are honored and grateful for the opportunity of observing the Universe from Maunakea, which has the cultural, historical, and natural significance in Hawaii.  T.D.B.~gratefully  acknowledges  support from the National   Aeronautics and Space Administration (NASA) under grant \#80NSSC18K0439 and from the Alfred P.~Sloan Foundation. We also thank Taichi Uyama for a Keck/NIRC2 dataset that was used to calibrate CHARIS astrometry.

\section*{Data Availability Statement}
The data underlying this article are available at \url{https://stars.naoj.org/}. The datasets underlying this article are reduced using the open-source python libraries: the \CHARIS Data Reduction Pipeline (\CHARIS DRP) \citep{Brandt_2017_CHARISPipeline} at \url{https://github.com/PrincetonUniversity/charis-dep} and pyKLIP \citep{pyKLIP_2015} at \url{https://pyklip.readthedocs.io/en/latest/}.

\label{lastpage}
\bibliographystyle{apj_eprint}
\bibliography{refs}

\end{document}